\documentclass[twocolumn]{aastex631}

\accepted{\today}

\shorttitle{JWST spectra of IR-faint WDs}
\shortauthors{Blouin et al.}
\begin{document}

\title{JWST Resolves CIA Features in White Dwarfs}

\author[0000-0002-9632-1436]{Simon Blouin}
\affiliation{Department of Physics and Astronomy, University of Victoria, Victoria, BC V8W 2Y2, Canada}

\author[0000-0001-6098-2235]{Mukremin Kilic} 
\affiliation{Homer L. Dodge Department of Physics and Astronomy, University of Oklahoma, Norman, OK 73019, USA}

\author[0000-0003-0475-9375]{Lo\"ic Albert}
\affiliation{D\'epartement de Physique, Universit\'e de Montr\'eal, Montr\'eal, QC H3C 3J7, Canada}
\affiliation{Institut Trottier de Recherche sur les exoplan\`etes, Universit\'e de Montréal, Canada}

\author[0000-0002-7898-6194]{Bianca Azartash-Namin}
\affiliation{Homer L. Dodge Department of Physics and Astronomy, University of Oklahoma, Norman, OK 73019, USA}

\author[0000-0003-4609-4500]{Patrick Dufour}
\affiliation{D\'epartement de Physique, Universit\'e de Montr\'eal, Montr\'eal, QC H3C 3J7, Canada}
\affiliation{Institut Trottier de Recherche sur les exoplan\`etes, Universit\'e de Montréal, Canada}



\begin{abstract}
Infrared-faint white dwarfs are cool white dwarfs exhibiting significant infrared flux deficits, most often attributed to collision-induced absorption (CIA) from H$_2$--He in mixed hydrogen--helium atmospheres. We present James Webb Space Telescope (JWST) near- and mid-infrared spectra of three such objects using NIRSpec (0.6--5.3$\,\mu$m) and MIRI (5--14$\,\mu$m): LHS~3250, WD~J1922+0233, and LHS~1126. Surprisingly, for LHS~3250, we detect no H$_2$--He CIA absorption at 2.4$\,\mu$m, instead observing an unexpected small flux bump at this wavelength. WD~J1922+0233 exhibits the anticipated strong absorption feature centered at 2.4$\,\mu$m, but with an unexpected narrow emission-like feature inside this absorption band. LHS~1126 shows no CIA features and follows a $\lambda^{-2}$ power law in the mid-infrared. LHS~1126's lack of CIA features suggests a very low hydrogen abundance, with its infrared flux depletion likely caused by He--He--He CIA. For LHS~3250 and WD~J1922+0233, the absence of a 1.2$\,\mu$m CIA feature in both stars argues against ultracool temperatures, supporting recent suggestions that infrared-faint white dwarfs are warmer and more massive than previously thought. This conclusion is further solidified by Keck near-infrared spectroscopy of seven additional objects. We explore possible explanations for the unexpected emission-like features in both stars, and temperature inversions above the photosphere emerge as a promising hypothesis. Such inversions may be common among the infrared-faint population, and since they significantly affect the infrared spectral energy distribution, this would impact their photometric fits. Further JWST observations are needed to confirm the prevalence of this phenomenon and guide the development of improved atmospheric models.
\end{abstract}


\section{Introduction} 
\label{sec:intro}
The atmosphere of a white dwarf becomes increasingly transparent as it ages and cools. As a result, the photosphere is located progressively deeper into the star and can reach liquid-like densities of a few ${\rm g}\,{\rm cm}^{-3}$ in the most extreme cases (i.e., an hydrogen-deficient atmosphere at an effective temperature of less than $\simeq 4000\,$K; \citealt{Saumon2022}). These conditions are unusual for stellar atmospheres, where ideal-gas conditions usually prevail (e.g., $\rho \sim 10^{-7}\,{\rm g}\,{\rm cm}^{-3}$ at the photosphere of the Sun). In cool white dwarf atmospheres ($T_{\rm eff} \lesssim 6000\,{\rm K}$), interactions between species affect the equation of state and the opacity of the gas. Modeling these non-ideal effects is challenging, but the increasing availability of quantum chemistry simulation codes and supercomputing resources has allowed significant progress \citep[e.g.,][]{Kowalski2007,Kowalski2010,Blouin2017,Blouin2018}. These advances have in turn led to a qualitative improvement in the quality of spectroscopic fits to the absorption features of cool white dwarfs \citep{Blouin2019WD2356,Blouin2019,Blouin2019DQpec}.

However, this progress still stops short of explaining the peculiar spectral energy distributions of infrared-faint white dwarfs.\footnote{IR-faint white dwarfs are often referred to as ultracool white dwarfs in the literature. However, \cite{Bergeron2022} suggest abandoning this term following their findings that these stars might not be so cool after all. We adhere to this convention in this work.} These objects suffer from significant absorption in the infrared (peaking at $2.4\,\mu$m), which results in surprisingly blue colors for their low temperatures ($T_{\rm eff} \lesssim 5000\,$K). This infrared absorption is attributed to collision-induced absorption (CIA) from H$_2-$X complexes in their mixed hydrogen--helium atmospheres \citep{Hansen1998,Jorgensen2000,Bergeron2002,Gianninas2015,Bergeron2022}. White dwarfs with pure-hydrogen atmospheres are not thought to significantly contribute to the observed IR-faint population, because CIA becomes dominant only at extremely low temperatures in those stars ($T_{\rm eff} \lesssim 4000\,$K) due to their lower photospheric densities (see Figure~16 of \citealt{Saumon2022}). Given the finite age of the Galactic disk, there are few white dwarfs that have had enough time to cool to such low temperatures. Similarly, white dwarfs with pure (or nearly pure) helium atmospheres lack the molecular hydrogen required for strong infrared absorption, although it has been suggested that He--He--He complexes constitute a significant source of CIA in some hydrogen-depleted white dwarfs \citep{Kowalski2014}.\footnote{CIA in a pure helium medium requires three-body collisions because interactions between two identical atoms are infrared inactive due to the lack of a net induced dipole moment. Three-body collisions break this symmetry, allowing for a temporary dipole moment and thus enabling infrared absorption.}

The existence of a population of cool white dwarfs with mixed hydrogen--helium atmospheres is most likely the result of convective mixing \citep{Rolland2018,Bedard2022,Bergeron2022}. As a white dwarf cools down, the superficial hydrogen convection zone deepens. If the hydrogen layer is thin enough ($M_{\rm H}/M_{\star} \lesssim 10^{-6}$), the bottom of this convection zone eventually reaches the much thicker helium layer underneath, thereby transforming a pure-hydrogen atmosphere into a mixed hydrogen--helium composition. As these convectively mixed objects cool down, they inevitably develop CIA and become IR faint.

\begin{figure}
\includegraphics[width=\columnwidth]{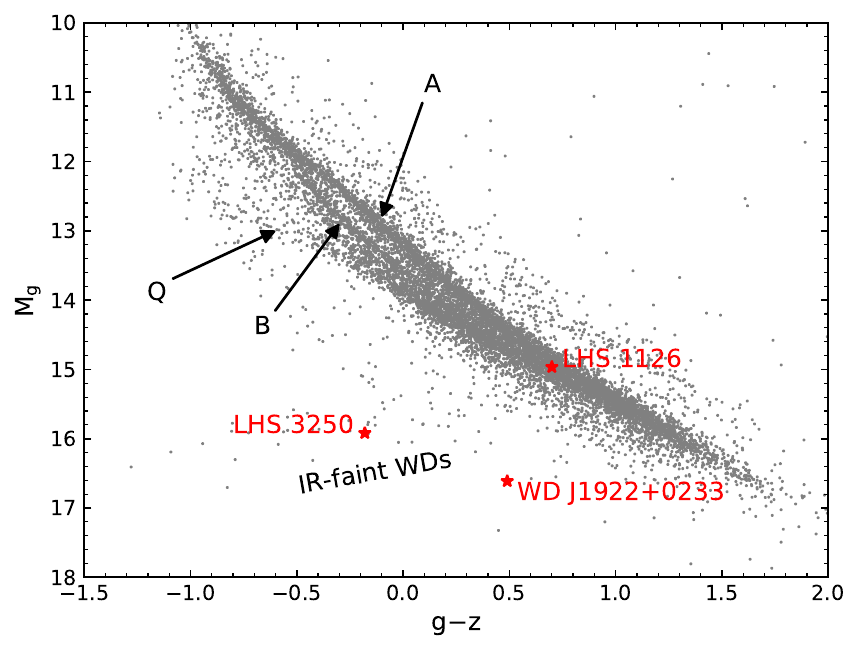}
\caption{Color--magnitude diagram of the Pan--STARRS white dwarfs within 100 pc of the Sun and with $10\sigma$ Gaia DR3 parallax measurements. Labels indicate the location of the A, B, Q, and IR-faint branches. The stars analyzed in this work are shown in red. The IR-faint branche starts at $M_{\rm g} \simeq 15$, $g-z \simeq 0.5$ and extends toward the bottom left corner of the figure.
\label{fig:HRD}}
\end{figure}

When CIA absorption is particularly intense, it can act beyond the infrared region and also reduce the emerging flux in the red optical \citep{Harris1999,Gates2004}, creating a distinct IR-faint sequence in the SDSS \citep{Kilic2020} or Pan--STARRS color--magnitude diagram (Figure~\ref{fig:HRD}). This sequence likely extends below the $M_{\rm g} \simeq 16.5$ cut-off visible in Figure~\ref{fig:HRD}, which is thought to be the result of Gaia's limiting magnitude \citep{Bergeron2022}. While there are now good explanations for the presence of the B and Q branches in this and similar color--magnitude diagrams \citep{Bergeron2019,Tremblay2019,Blouin2021,Camisassa2023,Blouin2023a,Blouin2023b,Bedard2024}, the exact origins and the properties of the stars belonging to the IR-faint branch remain uncertain.

To illustrate the extent of current uncertainties, consider the example of WD~J192206.20+023313.29 (hereafter WD~J1922+0233). This star was independently analysed by \cite{Elms2022} and \cite{Bergeron2022}. Different choices of input physics have led these two teams to converge to widely different atmospheric parameters. \cite{Elms2022} conclude that WD~J1922+0233 has a normal mass and an ultracool temperature ($0.57\,M_{\odot}$, $T_{\rm eff}=3340\,$K), while \cite{Bergeron2022} find a high-mass and not-so-cool solution ($1.07\,M_{\odot}$, $T_{\rm eff}=4440\,$K). It would be troubling enough to have such a high level of disagreement between two analyses, but equally concerning is that both teams' best solutions actually provide a poor match to the available photometric and spectroscopic data.\footnote{At least when ad hoc changes to the models' constitutive physics are not introduced.}

Because IR-faint white dwarfs only constitute a small fraction of the known white dwarf population (\citealt{Bergeron2022} were able to identify 105 IR-faint white dwarfs), it is tempting to ignore these outliers, considering them to be unimportant exceptions. However, we do not have that luxury. The possibility that IR-faint white dwarfs are among the coolest, oldest white dwarfs in our Galaxy makes them particularly interesting targets for age-dating applications. Yet, current uncertainties on their temperatures and masses make this impracticable. For example, there is a $\simeq 2\,$Gyr cooling age difference between the best-fit solutions of \cite{Elms2022} and \cite{Bergeron2022} for WD~J1922+0233. Furthermore, the atmospheres of some IR-faint white dwarfs are contaminated by planetary material \citep{Blouin2018b,Hollands2021,Kaiser2021,Elms2022}. This represents a unique window into the formation and evolution of planetary systems that arose during the Milky Way's infancy. Exploiting this opportunity requires reliable model atmospheres to accurately determine the chemical composition of the accreted planetary material. 

As alluded to above, \cite{Bergeron2022} recently suggested that most IR-faint white dwarfs are not as cool as previously believed. Earlier analyses of IR-faint samples found most IR-faint white dwarfs having $T_{\rm eff} \lesssim 4000\,$K, with many even being cooler than $3000\,$K \citep{Gianninas2015,Kilic2020}. There were at least two major reasons for doubting these ultracool temperatures. First, large white dwarf radii were needed to reconcile the observed luminosities with the ultracool temperatures. This led to very small white dwarf masses ($\sim 0.2\,M_{\odot}$). These can only be produced as the result of evolution in interacting binaries \citep{Brown2010}, since single-star evolution needs more than a Hubble time to produce such low-mass white dwarfs. The problem is that the existence of a sizeable population of extremely low-mass white dwarfs at ultracool temperatures is incompatible with the scarcity of such objects at higher temperatures \citep{Kilic2020}. Second, the agreement between the photometric data and model spectral energy distribution (SED) was very poor for most objects. \cite{Bergeron2022} presented new model atmospheres, which include more accurate helium opacities at high densities (\citealt{Iglesias2002}, see also \citealt{Kowalski2006,Blouin2018}). In these models, the increased transparency of helium makes the photosphere denser and H$_2-$He CIA stronger. This enabled \cite{Bergeron2022} to find much warmer solutions (most being in the $4000\,{\rm K} \leq T_{\rm eff} \leq 5000\,$K range) and therefore smaller radii and higher masses. In addition, the quality of the fits improved considerably.

While the analysis of \cite{Bergeron2022} apparently solves the two major issues that affected previous analyses of IR-faint samples, it generates new questions. First, the improved fits obtained by \cite{Bergeron2022} rely on the H$_2-$He CIA calculations of \cite{Jorgensen2000}. The more recent calculations of \cite{Abel2012} predict a significantly different absorption spectrum for H$_2-$He CIA, which does not lead to satisfying SED fits. This is perplexing because the \cite{Abel2012} CIA spectra are based on more detailed calculations than \cite{Jorgensen2000} and should a priori be considered more reliable. Second, the best-fit parameters of \cite{Bergeron2022} often lead to high-mass solutions ($\sim 1.0\,M_{\odot}$). The combination of relatively cool temperatures and high masses places these objects in the Debye cooling phase of white dwarf evolution \citep{Fontaine2001}. This is surprising because white dwarf cooling proceeds at an accelerated pace in this regime due to the rapidly falling heat capacity of the crystallized core, and therefore few objects are expected to be found in this phase.

To shed light on this issue and pave the way forward to reliable parameter determinations for these objects, we obtained infrared spectra of three IR-faint white dwarfs using the James Webb Space Telescope (JWST) Near-Infrared Spectrograph (NIRSpec, \citealt{Jakobsen2022}) and Mid-Infrared Instrument (MIRI, \citealt{Rieke2015}). For the first time, these observations resolve the CIA features of white dwarfs, providing new observational constraints that can be used to discriminate between models. We describe these observations in Section~\ref{sec:obs} and the models we use to analyze them in Section~\ref{sec:models}. Our analysis of the JWST spectra is then presented in Section~\ref{sec:fits}. As we will see, two of our JWST spectra exhibit an unexpected emission-like feature: we investigate its origin in Section~\ref{sec:bump}. Section~\ref{sec:keck} presents Keck near-infrared spectroscopic data for seven additional IR-faint white dwarfs that further support one of the key conclusions drawn from our analysis of the JWST spectra. We finally discuss the implications of our results for our understanding of IR-faint white dwarfs in Section~\ref{sec:discussion} along with our conclusions.

\section{JWST Observations}
\label{sec:obs}

\subsection{Choice of targets}
Our observing program (GO-3168, PI: Blouin) targeted three IR-faint white dwarfs: LHS~3250, WD~1922+0233, and LHS~1126. These objects were selected to provide the most useful constraints on IR-faint atmosphere models.

LHS 3250 is the prototypical IR-faint white dwarf \citep{Harris1999}. As the most studied object of its class, it was natural to include it in our sample. In addition, Spitzer photometry of LHS 3250 showed evidence of a flat or even increasing mid-infrared flux distribution \citep{Kilic2009}. This is not predicted by any model atmosphere, and JWST observations were deemed essential to resolve this discrepancy and potentially uncover new physics in IR-faint white dwarf atmospheres. 

WD J1922+0233 is a more recent addition to the IR-faint white dwarf catalogue \citep{Tremblay2020,Bergeron2022,Elms2022}. It is particularly interesting given the detection of metal lines in its optical spectrum. In theory, these features can provide constraints on the atmospheric density. As we have seen, recent analyses of WD J1922+0233 have led to conflicting results regarding its temperature and mass \citep{Bergeron2022,Elms2022}, making it an excellent test case for investigating current uncertainties in the modeling of IR-faint white dwarfs.

With an estimated effective temperature of 5200\,K \citep{Blouin2019,Bergeron2022}, LHS~1126 is one of the warmest known white dwarf exhibiting clear signs of CIA in the infrared. LHS~1126 has been observed with HST/FOS \citep{Wolff2002} and Spitzer/IRAC \citep{Kilic2006}. No model to date can simultaneously reproduce its entire SED \citep{Blouin2019}. While these discrepancies make LHS 1126 an ideal target for investigating the limitations in our understanding of opacity sources in cool white dwarf atmospheres, it remains unclear whether LHS 1126 is a unique oddity or representative of a broader class of objects. In fact, LHS~1126's optical spectrum also shows weak molecular carbon bands, which is unique among the IR-faint population.

\subsection{Observational setup and data reduction}
We obtained infrared spectra of our three targets using NIRSpec and MIRI. NIRSpec's low-resolution prism mode ($R \sim 100$) was used for the 0.6--5.3\,$\mu$m range, while MIRI's Low Resolution Spectrometer ($R \sim 100$) covered 5--14\,$\mu$m. Low-resolution spectroscopy is sufficient, as the CIA features are intrinsically broad. This setup provides continuous coverage across the entire 0.6--14\,$\mu$m range, ideal for tracing the overall spectral energy distribution and resolving broad molecular features.

We employed a 2-point dither strategy and chose integration times to achieve S/N $\geq 50$ at $4.5\,\mu$m (NIRSpec) and S/N $\geq 25$ at $8.0\,\mu$m (MIRI). Total integration times were 0.3\,h for LHS~1126, 2.4\,h for LHS 3250, and 4.7\,h for WD J1922+0233.

The data were reduced with the JWST calibration pipeline version 1.14.0
\citep{Bushouse2022}. Both the raw and fully reduced data were retrieved from the
Mikulski Archive for Space Telescopes (MAST). These data can be accessed via the DOI \dataset[10.17909/3sgc-yq02]{https://dx.doi.org/10.17909/3sgc-yq02}.

The pipeline-extracted NIRSpec prism spectra display a few spurious narrow absorption features. These features are found at different wavelengths in each dither position, and are not intrinsic to the source. We used the NIRSpec optimal spectral extraction notebook, but found no significant differences between the locally reduced data and the extracted data available from MAST. The spurious features are present in the optimally extracted spectra as well. For example, LHS~1126's spectrum shows a narrow dip at $4.72\,\mu$m in the first dither position and two narrow dips at 3.98 and $4.97\,\mu$m in the second dither position. These spurious features are not unique to our observations. For example, the NIRSpec data of the dusty white dwarf GD~362 obtained as part of the GO program 2919 also displays spurious narrow absorption features at various wavelengths in each dither position. Since our NIRSpec observations were obtained at only two dither positions, these spurious features make it to the final combined spectrum. Going forward, we recommend conducting NIRSpec observations with more than two dither positions (like in the case of GD~362) so that the spurious features can be rejected statistically.

\subsection{General description of the spectra}
\begin{figure}
\includegraphics[width=\columnwidth]{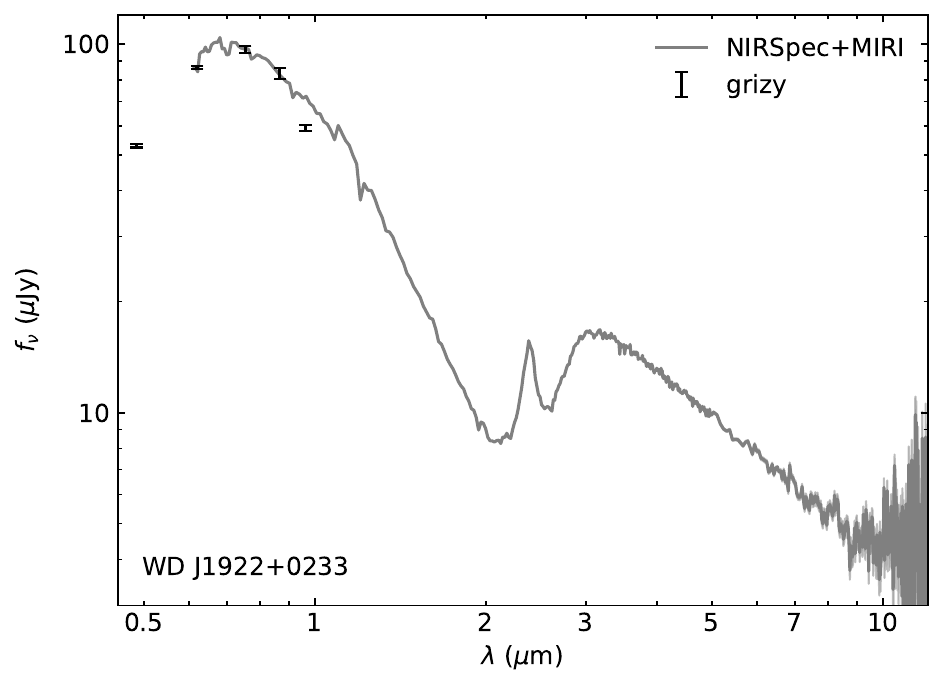}
\includegraphics[width=\columnwidth]{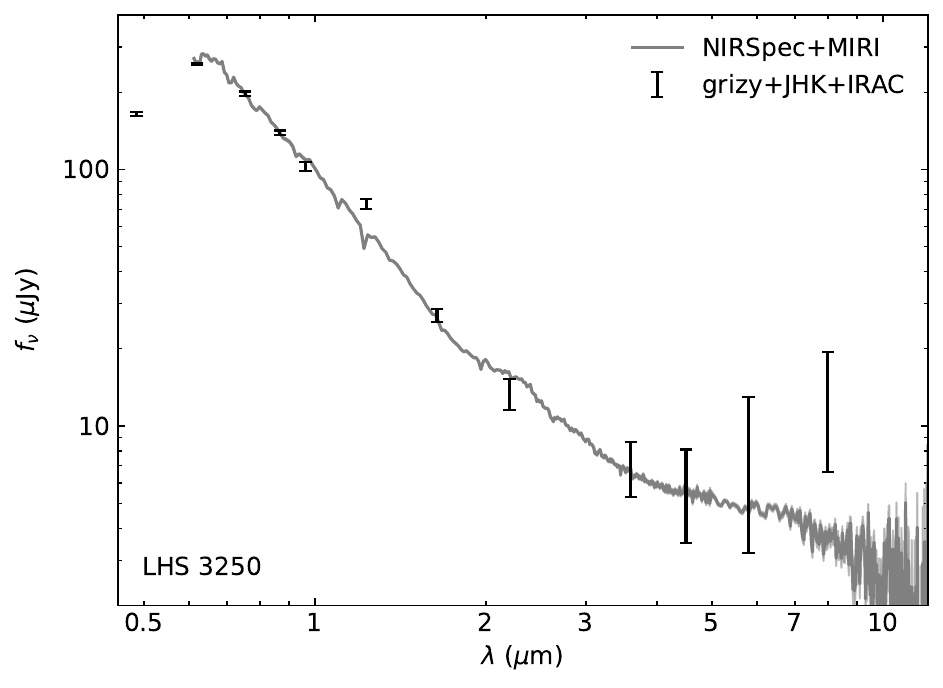}
\includegraphics[width=\columnwidth]{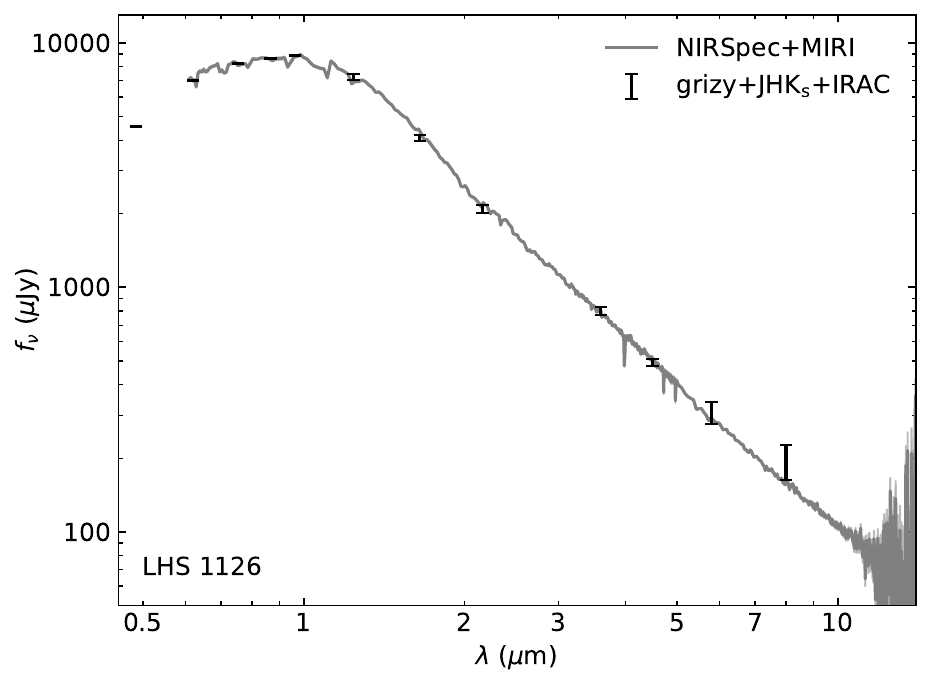}
\caption{NIRSpec and MIRI spectra of the three IR-faint white dwarfs observed as part of our program. Both spectra are merged at $5\,\mu$m. Photometry data from the literature is shown for comparison \citep{Harris1999,Kilic2006,Kilic2009,Skrutskie2006,Chambers2016}.}
\label{fig:obs}
\end{figure}

The combined NIRSpec and MIRI spectra for our three targets are presented in Figure \ref{fig:obs}. The two spectra were stitched at 5.0$\,\mu$m without any adjustment, and they merge seamlessly. For comparison, we also show photometry from other surveys and programs.

WD J1922+0233 exhibits by far the most striking spectrum of the three objects. The spectrum shows a strong absorption feature centered at 2.4$\,\mu$m, consistent with what we expect for H$_2$--He CIA (this is the fundamental H$_2$ vibrational band). Surprisingly, we observe a narrow emission-like feature inside this 2.4$\,\mu$m absorption band, which is not predicted by any existing model.

Against all expectations and model predictions, LHS~3250 shows no absorption feature at 2.4$\,\mu$m where CIA is expected to peak. Instead, we observe a small bump at that wavelength, which is not predicted by any current model. While the bump is subtle, it appears to be a real feature of LHS~3250's spectrum; we have no indications to the contrary. Unlike earlier tentative indications from relatively noisy Spitzer photometry \citep{Kilic2009}, we do not observe an increasing flux in the mid-IR for this object.

The mid-IR spectrum of LHS~1126 follows a $\lambda^{-2}$ power law with remarkable precision, consistent with the findings of \citet{Kilic2006} based on Spitzer photometry. Notably, similar to LHS~3250, there is no sign of a CIA absorption feature at 2.4$\,\mu$m, despite both H$_2$--H$_2$ and H$_2$--He CIA peaking at that wavelength.

The fact that both WD~J1922+0233 and LHS~3250 exhibit unexpected emission-like features at similar wavelengths is intriguing and suggests a possible common origin for these phenomena. The wavelength of 2.4$\,\mu$m corresponds to the $\Delta \nu = 1$ vibrational transition of the H$_2$ molecule. Therefore, there is a strong expectation, based on fundamental molecular physics, that H$_2-$He CIA is particularly opaque at that wavelength. This is also consistently predicted by all available CIA calculations \citep{Linsky1969,Borysow1989,Jorgensen2000,Abel2012,Blouin2017}. The emission-like features we observe at this wavelength thus represent a significant departure from basic expectations that cannot be easily explained by remaining opacity uncertainties alone. This discrepancy necessitates the exploration of alternative explanations, which will be the focus of Section~\ref{sec:bump}.

\begin{figure}
\includegraphics[width=\columnwidth]{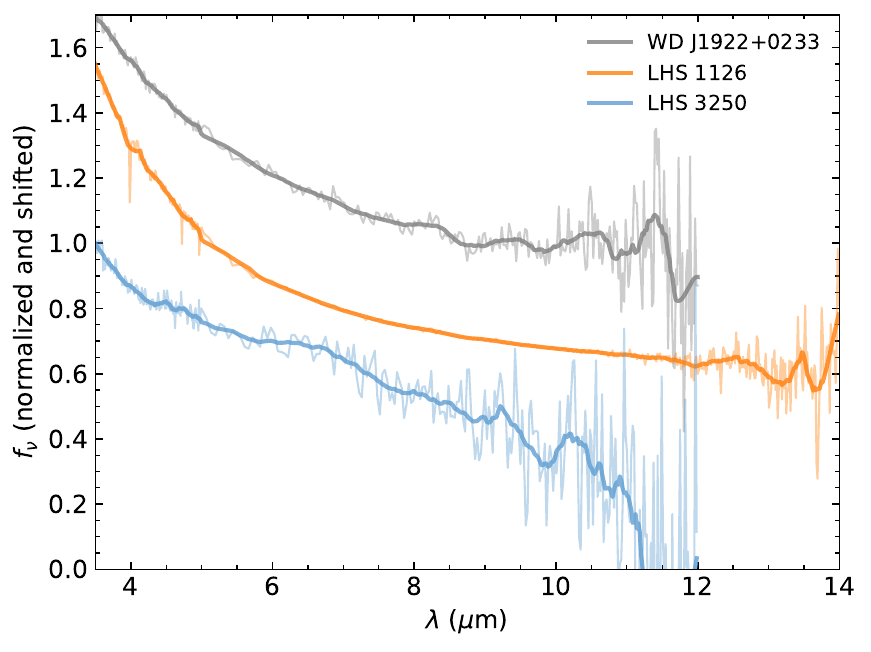}
\caption{JWST MIRI spectra of the three IR-faint white dwarfs in our sample, focused on the 3.5--14$\,\mu$m region. The spectra are normalized at 3.5$\,\mu$m and vertically offset for clarity. Thin lines show the original spectra, while thick lines represent smoothed versions using a Savitzky--Golay filter. No clear silicate emission features are detected in the 8--12\,$\mu$m range. For WD~J1922+0233 and LHS 3250, the spectra are cut off at 12\,$\mu$m due to poor signal-to-noise ratios at longer wavelengths.}
\label{fig:silicates}
\end{figure}

Finally, note that we do not detect any clear silicate emission features in the MIRI spectra of our three targets. Figure~\ref{fig:silicates} shows the 8--12\,$\mu$m region where silicate features have been observed in dusty white dwarfs \citep{Reach2005,Reach2009,Jura2009,Farihi2016,Swan2024}. While there are some fluctuations in the spectra, these appear consistent with the noise level of our observations.

\section{Model atmospheres}
\label{sec:models}
The model atmosphere calculations in this work are based on the code described in detail in \cite{Blouin2018,Blouin2018b}. This code incorporates several physical improvements for modeling cool, high-density white dwarf atmospheres, including a non-ideal equation of state \citep{Becker2014}, a refined treatment of helium ionization equilibrium following \cite{Kowalski2007}, high-density corrections to helium continuum opacities \citep{Iglesias2002}, and H$_2$--He CIA spectra from \cite{Abel2012}, with high-density corrections from \cite{Blouin2017}.

We use and expand a model atmosphere grid originally calculated by \cite{Blouin2019}. This grid spans effective temperatures of 3750\,K to 7000\,K in steps of 250 K, surface gravities $\log g = 7.0$ to 9.0 in steps of 0.5\,dex, and hydrogen-to-helium number ratios ranging from pure helium to pure hydrogen, with intermediate values of $\log {\rm H/He} = -5, -4.5, ..., 1.5, 2$. We extended the grid to lower temperatures (as cool as 3000\,K) and higher surface gravities (up to $\log g = 9.5$).

In addition to this extended grid, we calculated a second set of models based on the findings of \cite{Bergeron2022}. This second grid is identical to the first in all respects except for the adopted H$_2$--He CIA opacity. Instead of the \citet{Abel2012} CIA spectra, it uses the earlier calculations of \cite{Jorgensen2000}, including the density-dependent correction factor from \cite{Hare1958}. This dual-grid approach allows us to assess the impact of different CIA opacities on inferred white dwarf properties and to determine which best fit the JWST spectra.

To analyze the JWST spectroscopy, we employ a fitting procedure similar to the photometric method \citep{Bergeron2001}. Since the JWST spectra are well-calibrated in flux, we can directly fit $T_{\rm eff}$, the solid angle $\pi(R/D)^2$, and the hydrogen-to-helium ratio to the observed spectrum. This method involves minimizing the $\chi^2$ between the synthetic and observed spectra using the Levenberg--Marquardt algorithm. The solid angle, combined with the distance $D$ obtained from Gaia DR3 parallax measurements \citep{GaiaCollaboration2016,GaiaCollaboration2023}, allows us to determine the radius $R$ of the white dwarf. We use the mass--radius relation from evolutionary models \citep{Bedard2022} to derive the mass and surface gravity of the star from this $R$, thereby finding a self-consistent set of stellar parameters.

We do not attempt to provide error bars on our fitted parameters. The statistical uncertainties from the fitting procedure are very small and likely negligible compared to the systematic uncertainties stemming from the limitations in our understanding of the physics of these objects. The primary goal of this paper is not to provide precise characterization of individual stars, but rather to gain qualitative insights into the nature of IR-faint white dwarfs.

\section{Spectral fits}
\label{sec:fits}
\subsection{WD J1922+0233}
\label{sec:J1922}
\begin{figure}
\includegraphics[width=\columnwidth]{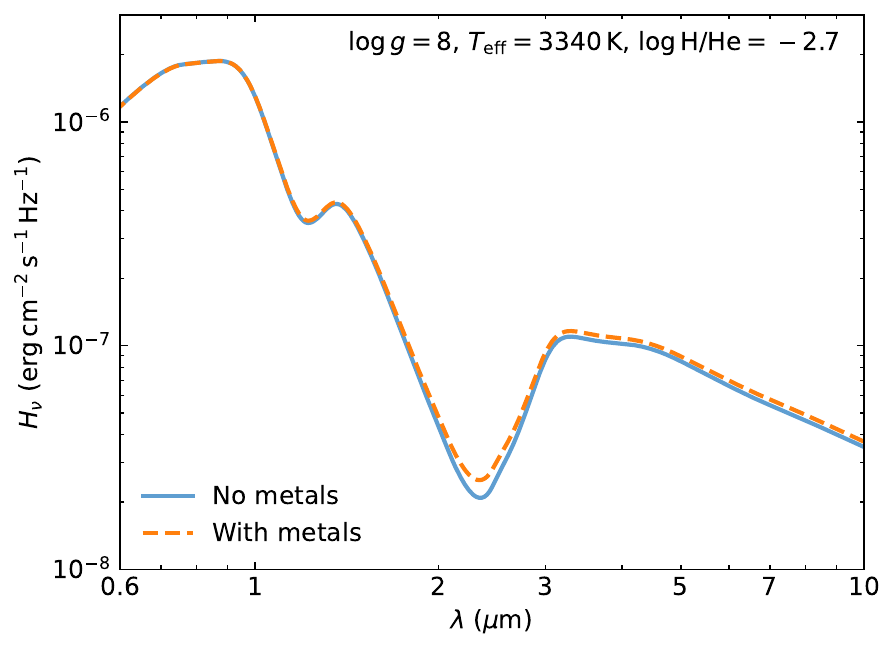}
\caption{Model spectra for a star with the same parameters as those obtained by \cite{Elms2022} for WD~J1922+0233. The solid line shows a model without metal pollution, while the dashed line corresponds to the case with metals ($\log\,{\rm Na/He}=-12.6$ and chondritic metal-to-metal abundance ratios for the other elements).
\label{fig:J1922-metal-impact}}
\end{figure}

Although WD J1922+0233 is known to be metal-polluted, we analyze it using metal-free models. This approach is justified because we expect the impact of metals on the overall spectral energy distribution to be minimal. This is illustrated in Figure \ref{fig:J1922-metal-impact}, where we compare a model with the metal abundance of \cite{Elms2022} to one without any metals. The difference between both models is insignificant. This aligns with \cite{Elms2022}, who reported that including metals in their fit of WD~J1922+0233 altered the synthetic photometry by less than 0.01\,mag. Adopting metal-free models significantly reduces the number of atmosphere calculations required, which is particularly advantageous given the convergence difficulties often encountered in this regime of physical parameters \citep{Bergeron1995}. These models frequently necessitate time-consuming manual interventions to achieve convergence. Moreover, as we will demonstrate, there are clearly more significant uncertainties in the models, such that the marginal effect of metals on the SED can be safely disregarded in the context of this analysis.

\begin{figure}
\includegraphics[width=\columnwidth]{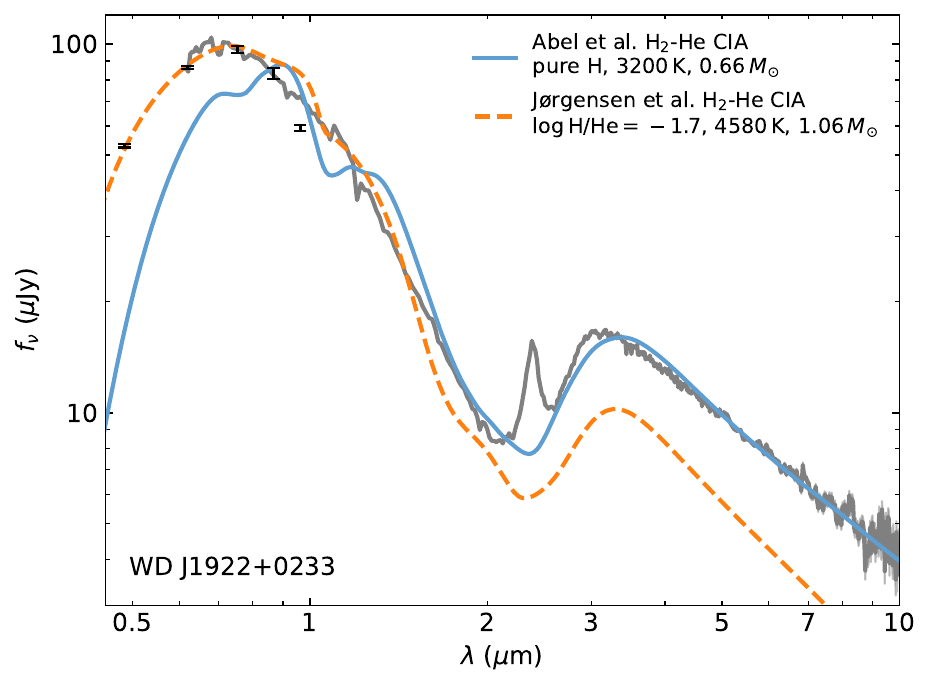}
\caption{Best-fit models to the JWST spectrum of WD~J1922+0233. The solid blue line displays the best fit for a grid of models using the \cite{Abel2012} opacities for H$_2-$He CIA, while the dashed orange line is for a grid of models using the \cite{Jorgensen2000} H$_2-$He CIA.
\label{fig:J1922-fits}}
\end{figure}

Figure \ref{fig:J1922-fits} presents the best-fit models for WD~J1922+0233 using our two model grids. Neither model reproduces the emission-like feature at 2.4$\,\mu$m, which is simply not predicted by standard atmosphere models. We will explore possible explanations for this feature in Section~\ref{sec:bump}.

The grid using \citeauthor{Abel2012} CIA opacities yields an ultracool solution (3200\,K), but interestingly, it converges on a pure hydrogen composition. Mixed hydrogen--helium solutions provide worse fits, meaning that the best-fit model shown here actually relies on the H$_2$--H$_2$ CIA opacities from \cite{Borysow2001} rather than the H$_2$--He opacities from \citeauthor{Abel2012} While this solution provides a reasonable fit to most of the infrared spectrum, it fails dramatically at wavelengths below $\simeq 1.5\,\mu$m. The good fit in the infrared comes at the expense of severely underpredicting the flux in the visible region. Moreover, it predicts a strong absorption band at $1.2\,\mu$m (the molecular hydrogen first overtone band) that is not observed in the NIRSpec spectrum. This $1.2\,\mu$m feature is even more pronounced in the best-fit solution of \citealt{Elms2022} (see their Figure~3).

In contrast, the solution obtained using the \citeauthor{Jorgensen2000} CIA opacities largely avoids these issues. It provides a good agreement in the optical region, and the $1.2\,\mu$m feature is much weaker due to the warmer temperature. While there is an offset in the infrared beyond $2\,\mu$m, the general shape of the spectrum is well reproduced. We will discuss a possible explanation for this offset in Section~\ref{sec:bump}. Note that this solution ($1.06\,M_{\odot}$, $T_{\rm eff} = 4580\,$K) is very similar to that found by \citealt{Bergeron2022} ($1.07\,M_{\odot}$, $T_{\rm eff} = 4440\,$K). This is despite differences in the models' input physics, notably the use of an ideal-gas equation of state in \cite{Bergeron2022} and a different treatment of helium pressure ionization.

\begin{figure}
\includegraphics[width=\columnwidth]{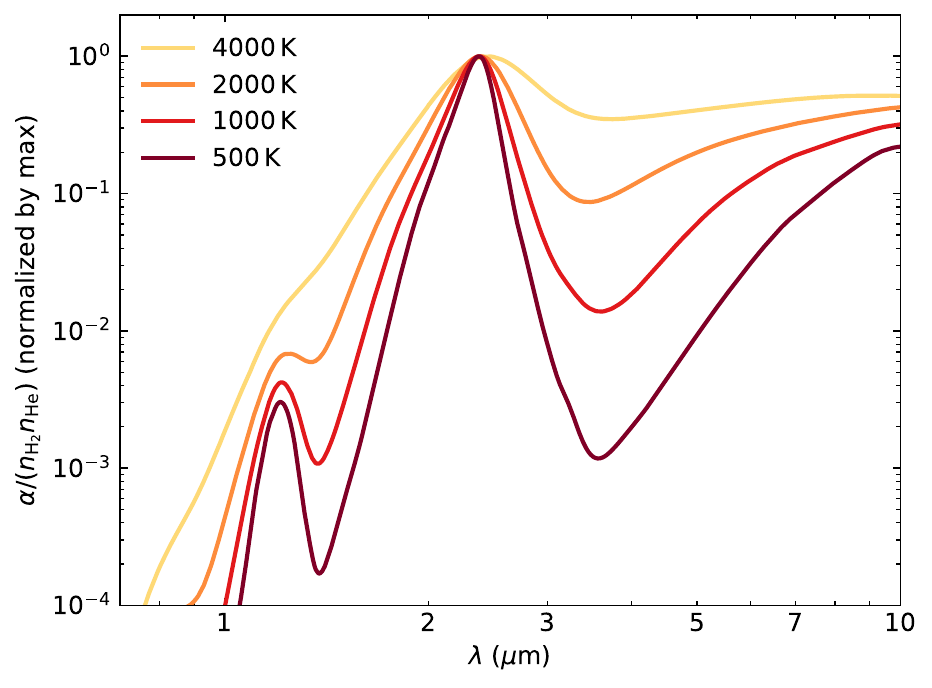}
\caption{H$_2-$He CIA spectra at different temperatures from \cite{Abel2012}. Each absorption spectrum is normalized at its maximum (without this normalization, the higher-temperature spectra would sit above the lower-temperature spectra). Note that the CIA-forming region of the atmosphere is typically much cooler than $T_{\rm eff}$ due to the high opacity at these wavelengths (Figure~\ref{fig:taunu}). This results in the emergent flux originating predominantly from the cooler, upper atmospheric layers.
\label{fig:CIA-T-dependence}}
\end{figure}

Overall, our analysis lends additional credence to the claim by \cite{Bergeron2022} that IR-faint white dwarfs are not as cool as previously thought. In particular, this conclusion is supported by the absence of a clear $1.2\,\mu$m feature in the JWST spectrum. At low temperatures, H$_2$--He and H$_2$--H$_2$ CIA spectra are expected to develop a pronounced and well-defined absorption feature at $1.2\,\mu$m (\citealt{Borysow1997,Borysow2001,Jorgensen2000,Abel2012}, see also Figure~\ref{fig:CIA-T-dependence}) due to reduced thermal broadening and enhanced dimer formation in a low-temperature gas \citep{Frommhold1993}. The absence of this feature, which should be particularly distinct if WD~J1922+0233 were indeed ultracool, suggests it is not so cool. Of course, we should caution that CIA spectra remain uncertain, but this pronounced $1.2\,\mu$m feature is a robust prediction of all available CIA calculations.

\subsection{LHS 3250}
\label{sec:LHS3250}
\begin{figure}
\includegraphics[width=\columnwidth]{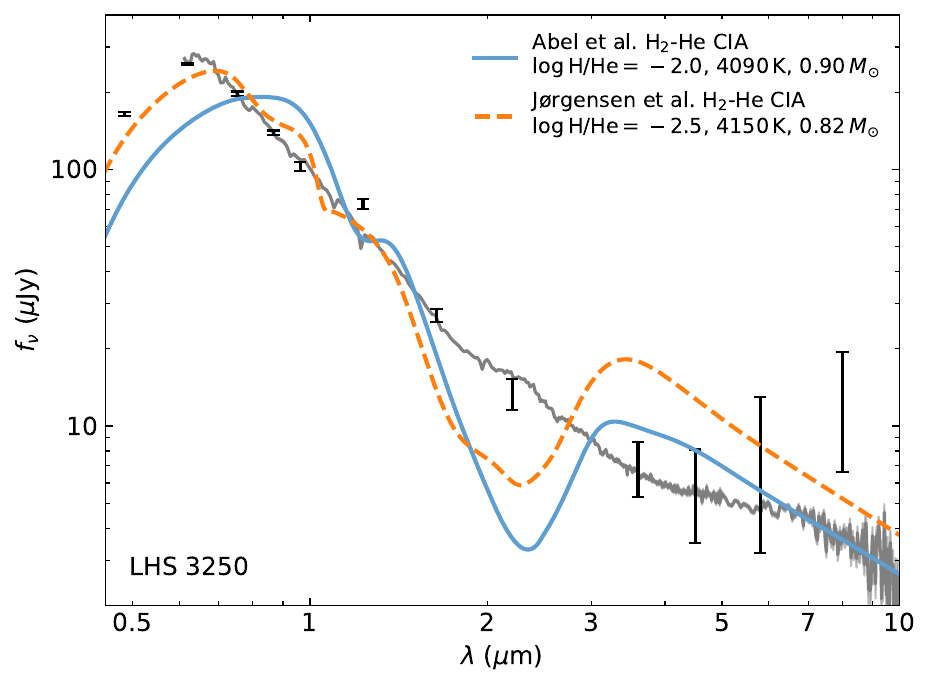}
\caption{Best-fit models to the JWST spectrum of LHS~3250. The solid blue line displays the best fit for a grid of models using the \cite{Abel2012} opacities for H$_2-$He CIA, while the dashed orange line is for a grid of models using the \cite{Jorgensen2000} H$_2-$He CIA.
\label{fig:LHS3250-fits}}
\end{figure}

Figure \ref{fig:LHS3250-fits} presents the best-fit models for LHS 3250 using our two model grids. Both solutions fail to reproduce the observed JWST spectrum due to their prediction of a strong H$_2$--He absorption band at 2.4$\,\mu$m that is surprisingly absent from the data. However, it is noteworthy that the small bump observed in the JWST spectrum aligns with the peak CIA absorption in both model fits. This intriguing feature will be further discussed in Section \ref{sec:bump}.

Despite the poor quality of the fits, both solutions yield similar stellar parameters. The grid using the \citeauthor{Abel2012} CIA opacities results in a mass of $0.90\,M_{\odot}$ and $T_{\rm eff} = 4090\,$K, while the \citeauthor{Jorgensen2000} grid yields $0.82\,M_{\odot}$ and $T_{\rm eff} = 4150\,$K. These temperatures are cooler than the solution found by \citealt{Bergeron2022} ($1.05\,M_{\odot}$, $T_{\rm eff} = 4990\,$K), but significantly warmer than earlier ultracool, low-mass solutions such as those of \citealt{Bergeron2002} ($0.23\,M_{\odot}$, $T_{\rm eff} = 3040\,$K) and \citealt{Gianninas2015} ($0.27\,M_{\odot}$, $T_{\rm eff} = 3060\,$K). However, given the poor quality of these fits, we do not consider these parameters to be reliable. Note also that the solution obtained using the \citeauthor{Abel2012} grid exhibits similar issues to those encountered with WD~J1922+0233 for the same model grid. It severely underpredicts the flux in the optical region and predicts a strong absorption feature at 1.2$\,\mu$m, in clear disagreement with the observations.

\subsection{LHS 1126}
\label{sec:LHS1126}
\begin{figure}
\includegraphics[width=\columnwidth]{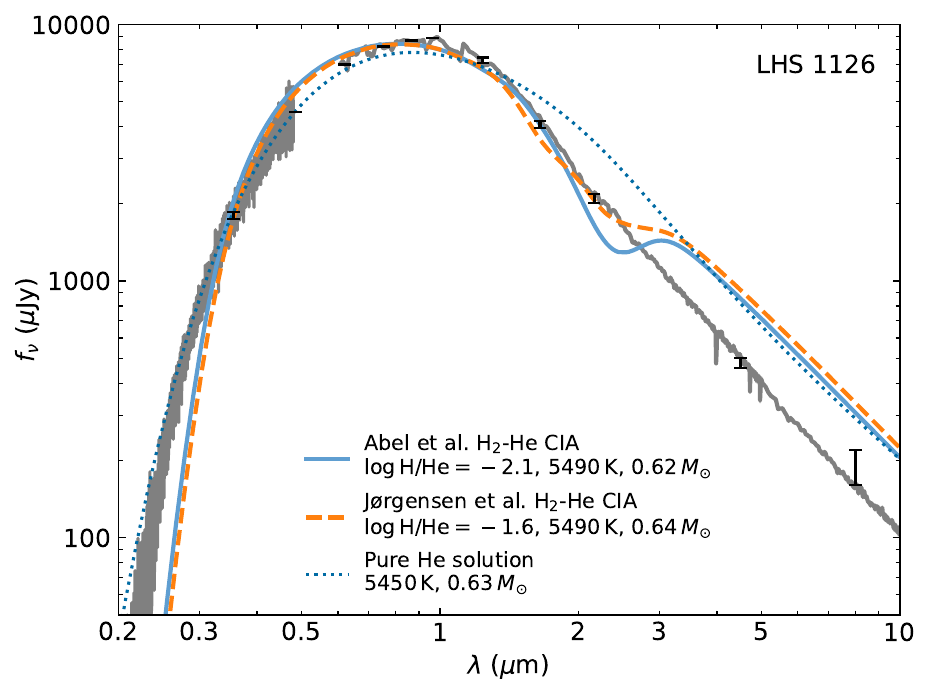}
\caption{Best-fit models for LHS~1126. The solid blue line displays the best-fit solution to the JWST data for a grid of models using the \cite{Abel2012} opacities for H$_2-$He CIA, while the dashed orange line is for a grid of models using the \cite{Jorgensen2000} H$_2-$He CIA. Also shown as a dotted dark blue line is the best-fit solution to the combined HST and JWST SED assuming a pure-helium atmosphere.
\label{fig:LHS1126-fits}}
\end{figure}

Figure \ref{fig:LHS1126-fits} presents our best-fit solutions to the JWST spectrum of LHS~1126. We also show its HST/FOS spectrum \citep{Wolff2002}, although it was not included in our fitting procedure to maintain consistency with our analysis of the two previous stars. Similar to LHS~3250, the fits are unsatisfactory due to the predicted 2.4$\,\mu$m feature in the hydrogen--helium solutions, which is absent in the JWST data. While the fit in the visible range is good, the ultraviolet flux is severely underestimated due to the red wing of the Ly-$\alpha$ line \citep[see also][]{Blouin2019}. 

The complete absence of any feature (absorption or emission-like) at 2.4$\,\mu$m strongly suggests a very low hydrogen abundance in LHS~1126's atmosphere. Given the precision of the JWST data, we found that any amount of hydrogen superior to $\log\,{\rm H/He}=-5$ would lead to a detectable absorption feature at 2.4$\,\mu$m, a result that applies both with the \citeauthor{Abel2012} and \citeauthor{Jorgensen2000} grids. This is consistent with the presence of (very weak) C$_2$ Swan bands in its optical spectrum, as cool DQ white dwarfs are known to have very low levels of hydrogen in their atmospheres. This is evidenced by the near-universal absence of CH features in such stars (\citealt{Blouin2019,Blouin2019DQpec}, Kilic et al. in prep.). Given these indications, we attempted to fit the full SED (taking into consideration both the HST and JWST data) using a hydrogen-free model atmosphere grid.\footnote{The very small carbon trace ($\log\,{\rm C/He}=-8.4$, \citealt{Blouin2019}) in LHS~1126's can be neglected. At the cool effective temperatures relevant to our analysis, it has no effect on the shape of the optical and infrared SED. Pressure-ionized helium is by far the main free electron contributor (see Figure~2 of \citealt{Blouin2023a}). This carbon abundance was also previously found to be too low to result in detectable C$_2-$He CIA \citep{Blouin2019}. The impact of carbon is limited to weak Swan bands and atomic absorption blueward of $0.2\,\mu$m. } While the overall fit is reasonable, there is insufficient absorption in the infrared to match the JWST data, despite the inclusion of He--He--He CIA from \cite{Kowalski2014} in our models. It is noteworthy that the three best-fit solutions shown in Figure~\ref{fig:LHS1126-fits} yield remarkably similar effective temperatures and masses.

\begin{figure}
\includegraphics[width=\columnwidth]{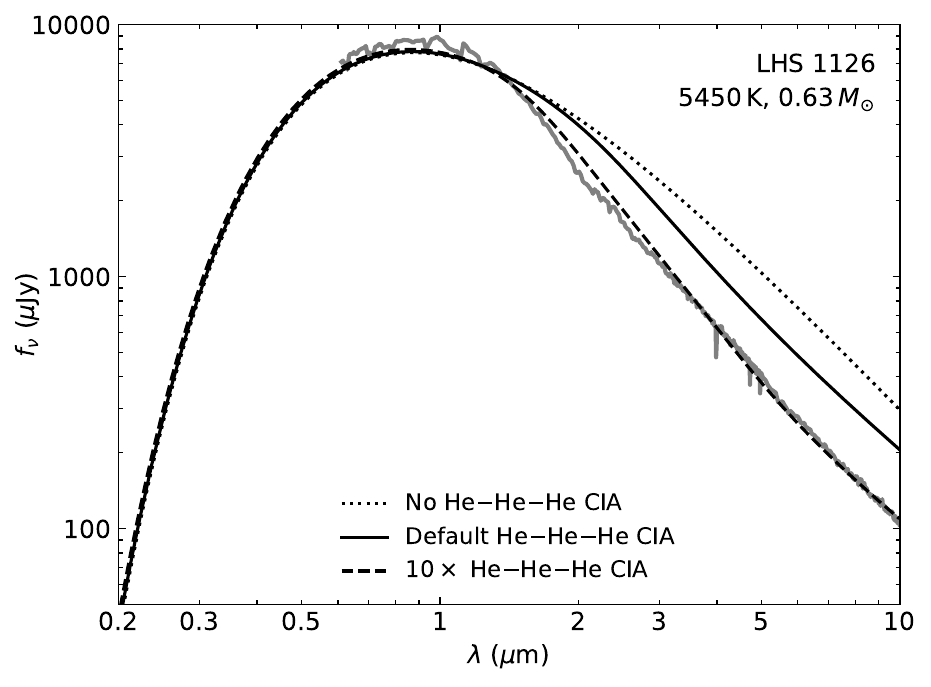}
\caption{Model spectra for a helium-atmosphere white dwarf with the same parameters as the best-fit pure-helium solution shown in Figure~\ref{fig:LHS1126-fits}. The solid line is the model obtained using the default He--He--He CIA of \cite{Kowalski2014}, as in the model shown as a dotted dark blue line in Figure~\ref{fig:LHS1126-fits}. The dotted and dashed lines illustrate the effect of decreasing or increasing this absorption source. The JWST data is shown for comparison. This supports the idea that an enhanced He--He--He CIA could explain LHS~1126's SED.
\label{fig:HeHeHe-SED}}
\end{figure}

The strength of He--He--He CIA is highly sensitive to density, scaling with the cube of the helium density. This implies that small adjustments to the non-ideal helium ionization at high densities, which controls the atmospheric density and remains highly uncertain \citep{Kowalski2007}, could significantly enhance He--He--He CIA. Adjustments to the He$^-$ free--free opacity could have a similar effect. Figure \ref{fig:HeHeHe-SED} illustrates how an enhanced He--He--He CIA could potentially explain the full SED of LHS~1126. Specifically, this figure demonstrates that a tenfold increase in He--He--He CIA would be sufficient to match the infrared flux level of LHS~1126, which corresponds to a density increase of only a factor of $\sim 2$. Such a change in the density structure of pure-helium model atmospheres is well within the uncertainties of our current understanding of non-ideal effects in high-density helium \citep{Saumon2022}.

\begin{figure}
\includegraphics[width=\columnwidth]{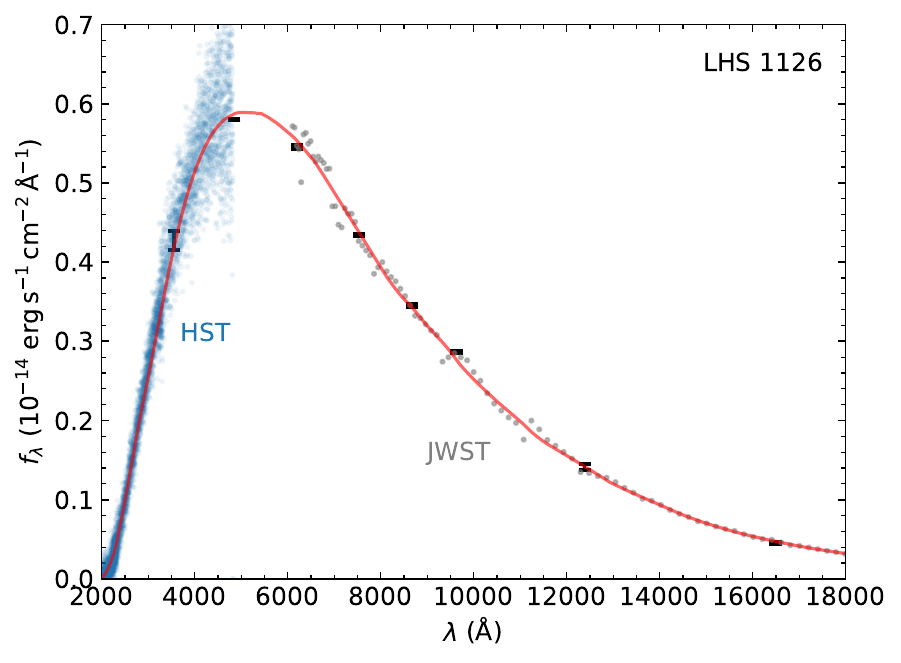}
\caption{Smoothed SED (red curve) used to calculate the total wavelength-integrated flux of LHS~1126. The HST and JWST data points used to build the smoothed SED are shown as blue and grey symbols. Archival optical and infrared photometry is also shown with black error bars.
\label{fig:LHS1126-full-SED}}
\end{figure}

\begin{figure}
\includegraphics[width=\columnwidth]{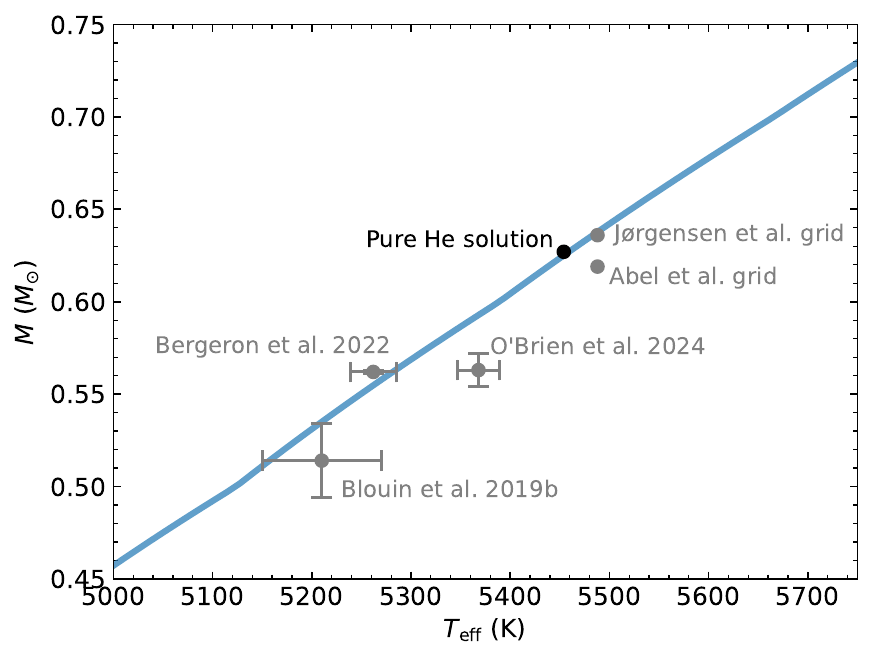}
\caption{The blue line marks the combinations of masses and effective temperatures that are compatible with the parallax and wavelength-integrated flux of LHS~1126. We have estimated the uncertainty to be comparable to the width of this line. The pure-helium solution of Figure~\ref{fig:LHS1126-fits} is indicated with a black circle. Gray symbols also report other solutions published in the literature \citep{Blouin2019,Bergeron2022,OBrien2024} as well as our mixed hydrogen--helium solutions (which we rejected).
\label{fig:LHS1126-constraints}}
\end{figure}

Given the extensive flux-calibrated data for LHS 1126 covering essentially its entire SED, we can also establish constraints on its atmospheric parameters that are independent of model atmospheres. We first performed a simple smoothing of the available HST+JWST data with a Savitzky--Golay filter (Figure \ref{fig:LHS1126-full-SED}) to obtain a complete SED, which we then integrated. This integral is directly related to $T_{\rm eff}$ and the star's radius in a way that is completely independent of model atmospheres:
\begin{equation}
\int f_{\lambda} d\lambda = \sigma T_{\rm eff}^4 \frac{R^2}{D^2},
\end{equation}
where $\sigma$ is the Stefan--Boltzmann constant. Using a standard mass--radius relation \citep{Bedard2022} and the Gaia-based distance, we derived a well-defined relation for the possible mass--$T_{\rm eff}$ values of LHS~1126 (Figure \ref{fig:LHS1126-constraints}). This figure shows that our pure-He solution of Figure~\ref{fig:LHS1126-fits} perfectly matches the model-independent constraint. While this agreement is not surprising, given that we used the same HST and JWST data in our model fitting process, it is nonetheless a welcome reassurance considering the high uncertainties surrounding IR-faint models.

\section{On the origin of the emission-like features}
\label{sec:bump}
We now explore possible causes for the emission-like feature at 2.4$\,\mu$m detected in LHS~3250 and WD~J1922+0233. We exclude LHS~1126 from this discussion, as the absence of any feature in its infrared spectrum (and therefore extremely low hydrogen abundance) places it in a distinct category from the other two IR-faint white dwarfs.

\subsection{Density distortion effects}
\label{sec:density_distortion}
\begin{figure}
\includegraphics[width=\columnwidth]{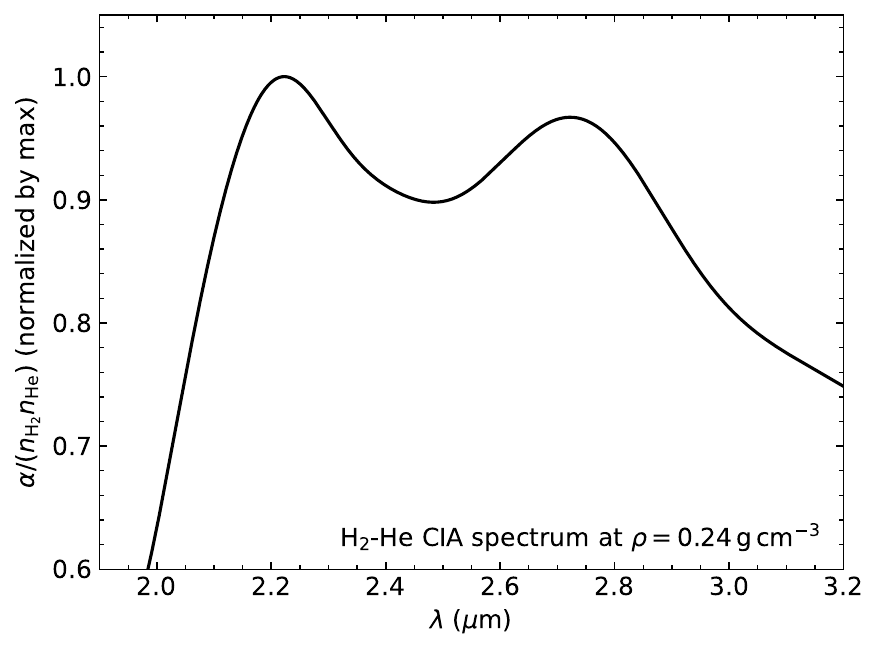}
\caption{H$_2-$He CIA spectrum at $\rho=0.24\,{\rm g}\,{\rm cm}^{-3}$ according to \cite{Blouin2017}. Note the split of the 2.4$\,\mu$m band. This particular spectrum was obtained at $T=5000\,$K.
\label{fig:CIA-distort}}
\end{figure}

Based on density functional theory molecular dynamics simulations, \cite{Blouin2017} predict that at densities above 0.1\,g\,cm$^{-3}$, the 2.4$\,\mu$m H$_2$--He CIA band can split into two components. The resulting CIA spectrum then exhibits a local minimum between two absorption peaks (Figure \ref{fig:CIA-distort}). At first glance, this effect appears similar to what we observe in the infrared spectrum of WD~J1922+0233 (Figure~\ref{fig:J1922-fits}), where the emission-like feature could be interpreted as the local minimum of the CIA spectrum. 

However, there are significant problems with this interpretation. First, this effect is already included in our model grid, and yet it does not appear in our best-fit models. We also fail to see this effect in much cooler models (e.g., for the atmospheric parameters of WD~J1922+0233 determined by \citealt{Elms2022}), as shown in Figure~12 of \cite{Blouin2017}. This is primarily because in the CIA-forming regions of the atmosphere, which are well above the photosphere due to the strong opacity at wavelengths affected by CIA, the density is simply too low for these CIA distortion effects to significantly impact the emerging spectrum. One might argue that this could indicate that the density is severely underestimated in these cool hydrogen--helium models. However, there are already clear indications that the density might actually be overestimated. Using the same models as those used here, \cite{Bergeron2022} found it impossible to account for the narrowness of WD~J1922+0233's Na absorption line (see their Figure~11). The predicted Na feature is much too wide, suggesting that the density in the line-forming region of the atmosphere is significantly lower than predicted by the models. \cite{Elms2022} encountered a similar problem and chose to arbitrarily reduce the broadening constant of the offending lines by a factor of 100.

A second issue is that this interpretation fails to explain the spectrum of LHS~3250. The density distortion effect only predicts a reduction in absorption at the center of the 2.4$\,\mu$m band, not a complete elimination of absorption across the band or the presence of an emission-like feature. An absorption profile like that shown in Figure \ref{fig:CIA-distort}, even with significant distortion, would still result in net absorption across the entire band, which is inconsistent with the observations of LHS~3250. These issues strongly suggest that density distortion effects alone are insufficient to explain the observed spectral features in WD~J1922+0233 and LHS~3250.

\subsection{Temperature inversion above the photosphere}
\label{sec:Tinv}
\begin{figure}
\includegraphics[width=\columnwidth]{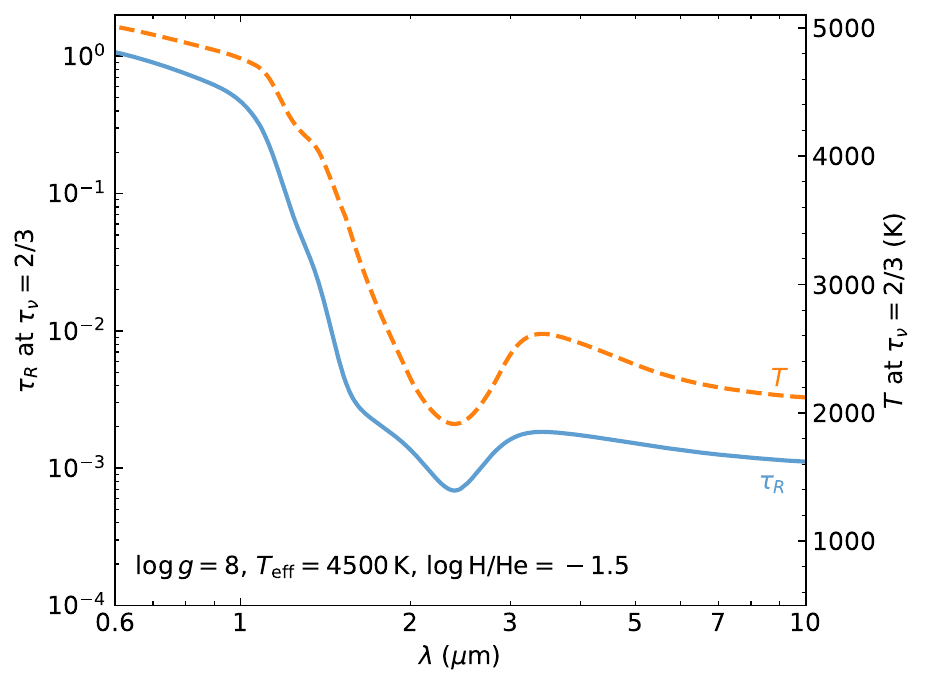}
\caption{Rosseland mean optical depth at which $\tau_{\nu}=2/3$ as a function of wavelength (solid line, left axis) and temperature at that same depth (dashed line, right axis). Note how the strong CIA opacity implies that the infrared portion of the spectrum is formed much higher in the atmosphere than the photosphere.
\label{fig:taunu}}
\end{figure}

A temperature inversion in the upper atmosphere of WD~J1922+0233 and LHS~3250 could potentially explain their emission-like feature. Due to the high opacity at CIA-forming wavelengths, we probe increasingly higher atmospheric layers as we approach peak CIA absorption at 2.4$\,\mu$m. Figure \ref{fig:taunu} illustrates this effect, showing in blue the Rosseland optical depth from which the typical photon emerges (i.e., where $\tau_{\nu}=2/3$) for a model with parameters typical of hydrogen--helium atmosphere IR-faint white dwarfs. At 2.4$\,\mu$m, the typical photon emerges from a region where $\tau_R \lesssim 10^{-3}$, well above the photosphere (where $\tau_R=2/3$ by definition). Under standard conditions, where temperature decreases from the photosphere outward, the temperature in this region is lower than at the photosphere (dashed orange line in Figure~\ref{fig:taunu}). However, if the temperature happens to be higher in these upper atmospheric levels, the star could appear brighter at these wavelengths.

\begin{figure}
\includegraphics[width=\columnwidth]{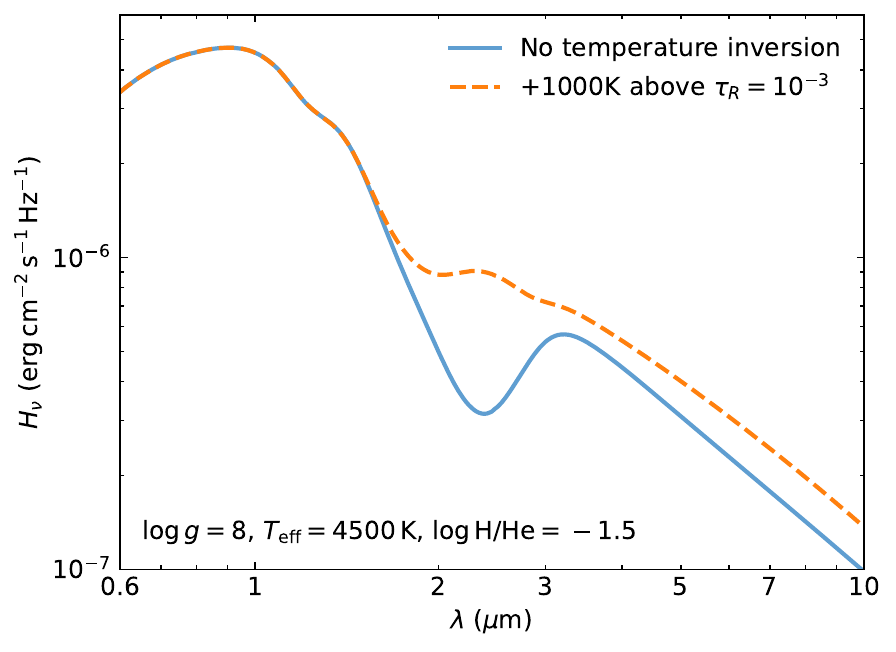}
\caption{Model spectrum of an IR-faint white dwarf based on a thermodynamic structure found using the standard linearisation technique to reach radiative equilibrium (solid line) and model spectrum for the same star but with an ad hoc 1000\,K increase in the temperature profile above $\tau_R = 10^{-3}$ (dashed line).
\label{fig:inversion_spectra1}}
\end{figure}

Figure \ref{fig:inversion_spectra1} demonstrates this effect. We artificially increased the temperature of the uppermost layers ($\tau_R < 10^{-3}$) by 1000\,K. This does not affect the flux below 1.5$\,\mu$m, where the continuum forms deeper than $\tau_R = 10^{-3}$. However, it increases the flux where CIA is strong, particularly at its peak. This results in enhanced thermal emission at 2.4$\,\mu$m, creating a small flux bump not too different to that observed in LHS~3250 (Figure~\ref{fig:LHS3250-fits}). We stress that this 1000\,K boost is entirely ad hoc, serving solely as a proof of concept. Possible explanations for such a temperature inversion are discussed in Section~\ref{sec:causes_Tinv}.

% \begin{figure}
% \includegraphics[width=\columnwidth]{figures/inversion-Ttau.pdf}
% \caption{Temperature at the atmospheric level where $\tau_{\nu}=2/3$. The solid line corresponds to a standard model structure, while the dashed line displays a case where an ad hoc 1000\,K increase in the temperature profile above $\tau_R = 10^{-3}$ has been applied (as in Figure~\ref{fig:inversion_spectra1}). Note how a local maximum appears at $2.4\,\mu$m, where the CIA opacity is strongest. This is responsible for the thermal emission that erases the $2.4\,\mu$m absorption band in Figure~\ref{fig:inversion_spectra1}.
% \label{fig:inversion-Ttau}}
% \end{figure}

\begin{figure}
\includegraphics[width=\columnwidth]{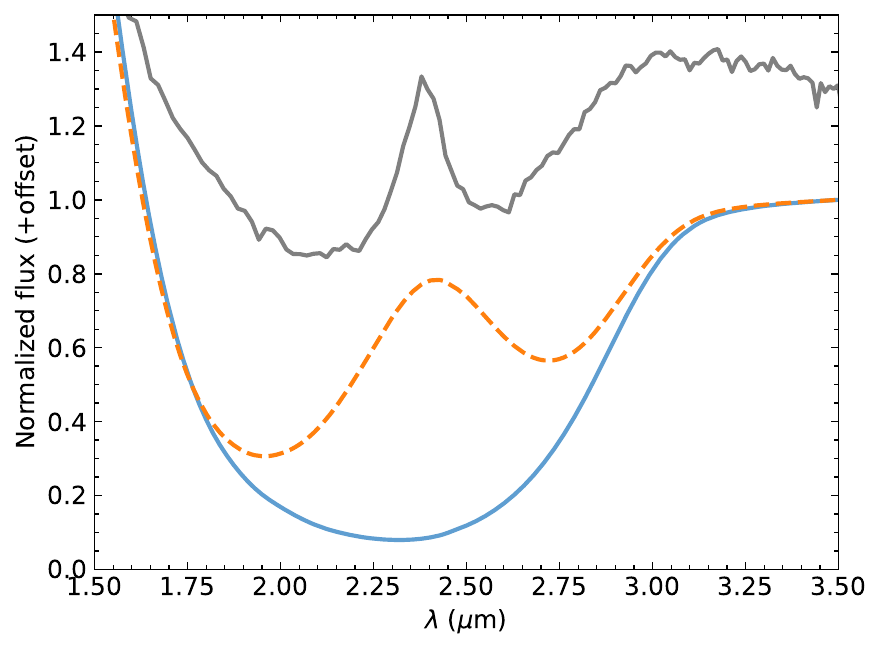}
\includegraphics[width=\columnwidth]{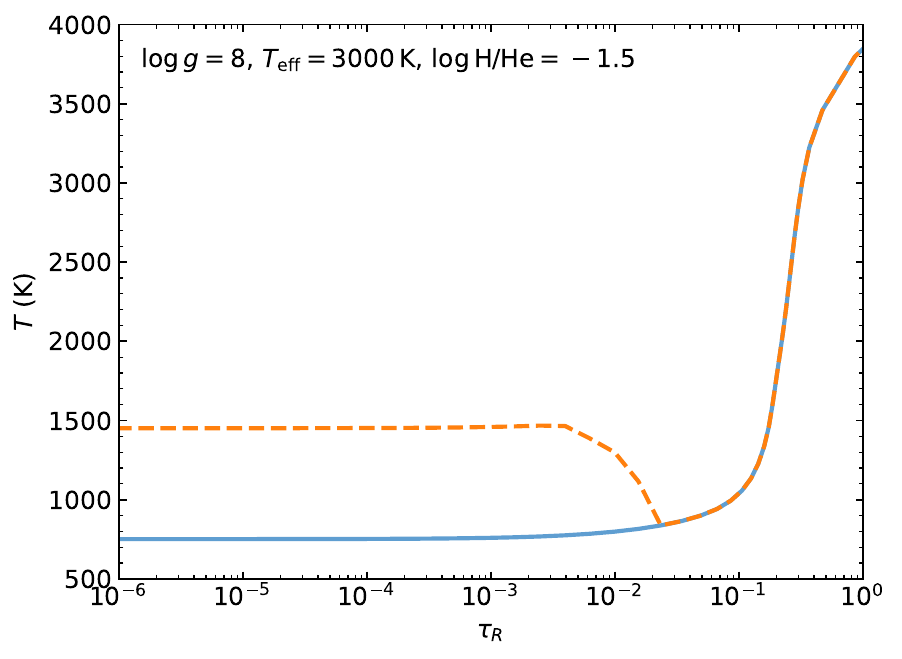}
\caption{\textit{Top}: JWST spectrum of WD~J1922+0233 (grey) in the H$_2$ fundamental band region, compared to a standard ultracool model ($T_{\rm eff}=3000\,{\rm K}$, $\log\,{\rm H/He}=-1.5$; blue solid line) and the same model with an ad hoc temperature inversion in the upper atmosphere (orange dashed line). All three spectra were normalized at $3.5\,\mu$m, and the JWST spectrum was vertically shifted for clarity. \textit{Bottom}: Temperature structures underlying the two model spectra shown in the top panel.
\label{fig:inversion-cool}}
\end{figure}

Reproducing the narrow emission-like feature seen in WD~J1922+0233's spectrum with a temperature inversion has proved more challenging. A key difficulty is that increasing the temperature broadens the 2.4$\,\mu$m feature (Figure~\ref{fig:CIA-T-dependence}), conflicting with the narrow feature observed in WD~J1922+0233. Consequently, we could only obtain an emission-like feature at 2.4$\,\mu$m resembling WD~J1922+0233's spectrum by assuming a much lower $T_{\rm eff}$ than found in our earlier analysis (Figure~\ref{fig:inversion-cool}). Reducing $T_{\rm eff}$ results in lower temperatures in the region above the photosphere, which in turn allows us to introduce a temperature inversion while still maintaining temperatures low enough to preserve a narrow CIA feature. 

The temperature inversion hypothesis appears particularly promising for explaining the spectral features of LHS~3250, but presents more challenges when applied to WD~1922+0233 due to the narrowness of the observed emission-like feature. However, it would be premature to discard the temperature inversion idea for WD J1922+0233 based on this issue alone. 

First, Occam's razor suggests we should prefer a common explanation for the emission-like features of both LHS~3250 and WD~J1922+0233, rather than invoking separate mechanisms for each star. 

\begin{figure}
\includegraphics[width=\columnwidth]{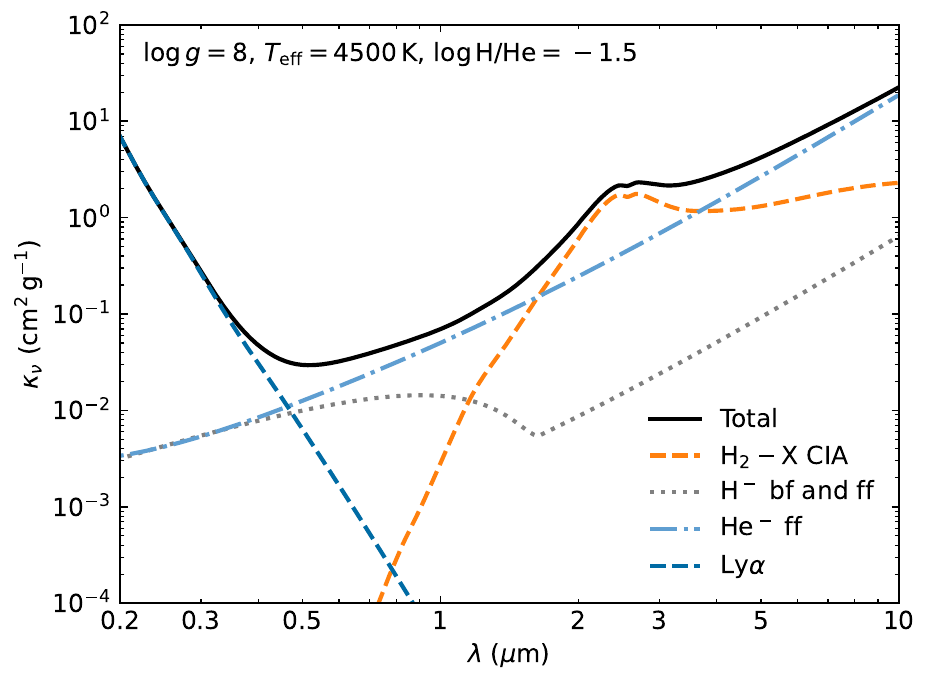}
\caption{Main contributions to the total opacity at the photosphere ($\tau_R=2/3$) of a typical IR-faint white dwarf. Note the small split in the 2.4$\,\mu$m CIA band opacity. Consistent with our discussion in Section~\ref{sec:density_distortion}, this split is far too weak to significantly affect the emerging spectrum.
\label{fig:contribopac}}
\end{figure}

Second, the opacity sources that shape the temperature structure of cool hydrogen--helium IR-faint white dwarf atmospheres (Figure~\ref{fig:contribopac}) are notoriously uncertain \citep{McWilliams2015,Saumon2022}. The low-mass problem identified in the Gaia data for cool white dwarfs probably points to significant problems with opacities for pure hydrogen atmospheres at low temperatures \citep{Caron2023,OBrien2024}. These issues are likely exacerbated in hydrogen--helium atmospheres due to their higher densities and more complex opacity physics. For a model with $T_{\rm eff} = 4500\,$K, $\log g = 8$, and $\log {\rm H/He} = -1.5$, we find that changing the He$^-$ free-free opacity or the H$^-$ opacities by a factor of 10 results in a 1000\,K change in the temperature at $\tau_R = 10^{-3}$. In addition, three-dimensional simulations of pure-hydrogen atmospheres have revealed a significant reduction of the temperature in the uppermost layers due to overshooting motions that force the entropy gradient in the stable layers above the convection zone to approach a near-adiabatic structure \citep{Tremblay2013}. Naturally, a similar effect can be expected to impact hydrogen--helium atmospheres. For these reasons, the temperature in the uppermost layers of our IR-faint model atmospheres may be overestimated by a non-negligible margin. This would in turn allow for a more significant temperature inversion without reaching temperatures that would overly broaden the CIA features.

Third, non-local thermodynamic equilibrium (non-LTE) effects in the uppermost region of the atmosphere cannot be ruled out. The models used in this work assume LTE, and a definitive assessment of non-LTE effects would require calculations using a non-LTE atmosphere code that incorporates all relevant microphysics for cool hydrogen--helium atmosphere white dwarfs. Such a code is not currently available. If present, non-LTE effects could potentially result in H$_2-$He collision-induced emission (CIE; \citealt{Frommhold1993}), which might contribute to the narrow feature observed in WD~J1922+0233 at $2.4\,\mu$m.

Finally, as discussed above, WD~J1922+0233 also displays a surprisingly narrow Na line \citep{Bergeron2022,Elms2022}. This narrow feature is particularly perplexing when compared to other very cool DZ stars, which typically show very broad Na features that are satisfactorily reproduced by existing models \citep{Blouin2019,Kaiser2021}. The fact that a star with ostensibly similar parameters behaves so differently cannot be easily explained by systematic problems with the model atmospheres' microphysics, as these would affect all cool DZ stars similarly. This suggests that the peculiarity must be more specific to WD~J1922+0233 itself. In this context, invoking a peculiar temperature stratification for WD~J1922+0233 becomes more plausible.

Given these considerations, while challenges remain in fully explaining the narrow emission-like feature in WD~J1922+0233, we view the temperature inversion hypothesis as a promising avenue for understanding the infrared spectra of both LHS~3250 and WD~J1922+0233.

\subsection{Possible causes of the temperature inversion}
\label{sec:causes_Tinv}
If a temperature inversion in the upper atmosphere is indeed responsible for the emission-like features observed in LHS~3250 and WD~J1922+0233, then what is causing it? We explore two main scenarios.

One potential explanation draws parallels with DAe white dwarfs, which exhibit Balmer lines in absorption with emission cores at their centers \citep{Elms2023}. These objects are thought to be related to the more numerous DAHe white dwarfs, where Zeeman splitting is also detected \citep{Greenstein1985,Manser2023}. The spectral features of DA(H)e stars are likely explained by an intrinsic temperature inversion (chromosphere) supported by the white dwarf's magnetic field \citep{Walters2021,Lanza2024}. A similar mechanism could be at work in IR-faint white dwarfs, with the emission-like feature in the H$_2-$He CIA band analogous to the Balmer line emission cores in DA(H)e stars. This scenario would require the presence of a magnetic field in LHS~3250 and WD~J1922+0233. While no magnetic field has been detected in these objects, DAe white dwarfs also lack detectable fields, yet one is likely required to explain their emission features.

Recent evidence suggests an increased occurrence of magnetic fields in cool white dwarfs \citep{Bagnulo2022}, although the picture below $T_{\rm eff}=5000\,$K is less clear due to the need for spectropolarimetry to detect magnetic fields in the absence of Balmer lines \citep{Berdyugin2022,Berdyugin2024}. It is thus a priori plausible that a significant fraction of IR-faint white dwarfs possess magnetic fields strong enough to support a chromosphere. \cite{Harris1999} presented spectropolarimetry of LHS~3250 and did not detect a magnetic field, but it is still unclear how strong the field needs to be to create a temperature inversion.

\begin{figure}
\includegraphics[width=\columnwidth]{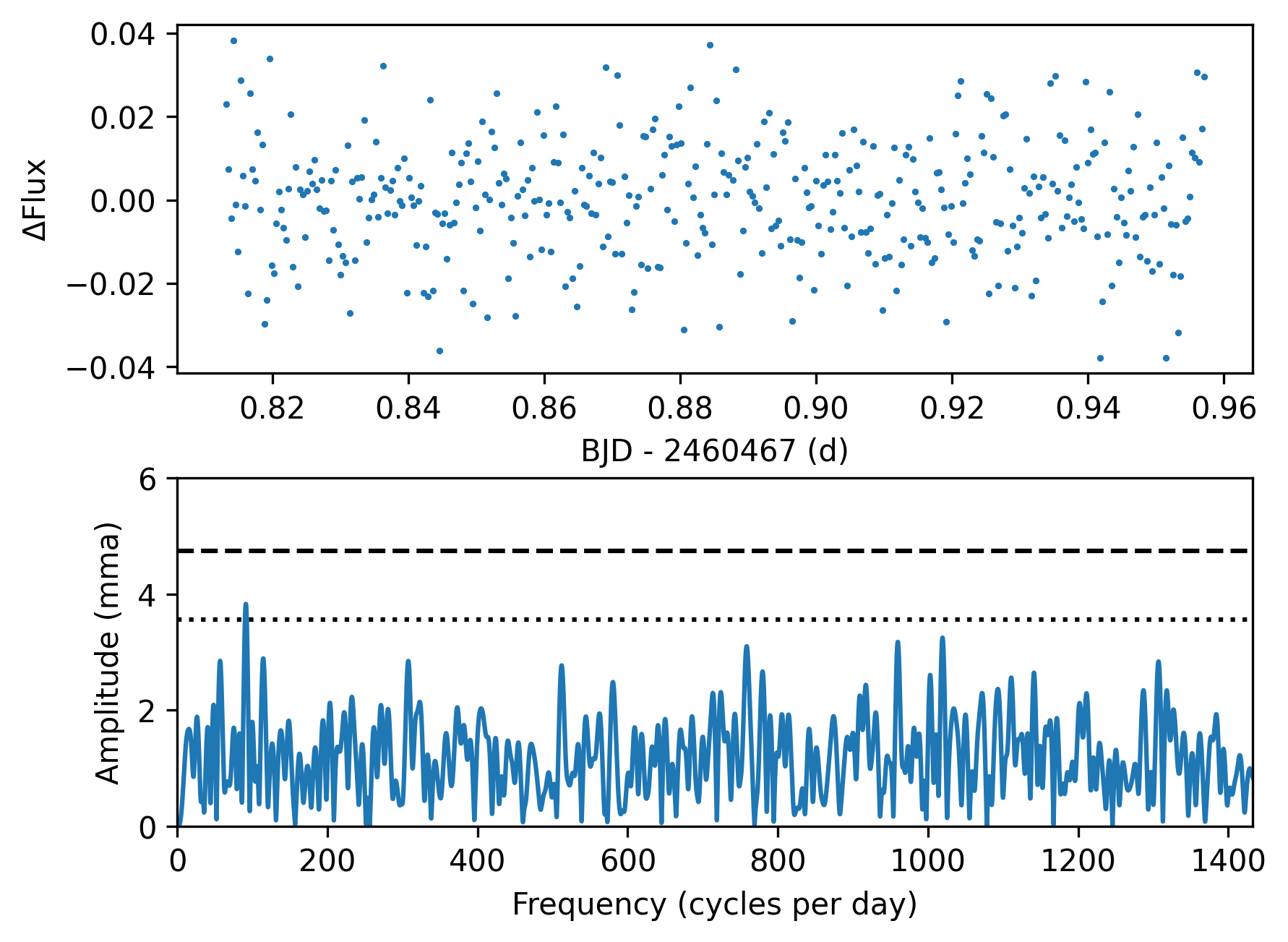}
\caption{High-speed photometry of WD~J1922+0233 obtained with the APO 3.5\,m telescope. The top panel shows the light curve, and the bottom panel displays its Fourier transform plotted up to the Nyquist frequency.  The dotted and dashed lines show the $3\langle A \rangle$ and $4\langle A \rangle$ levels (3.6 and 4.7\,mma, respectively), where $\langle A \rangle$ is the average amplitude in the Fourier transform. There is no evidence of any significant photometric variability in the APO data for WD~J1922+0233.
\label{fig:J1922-photometry}}
\end{figure}

To further investigate this scenario, we searched for photometric variability in LHS~3250 and WD~J1922+0233, as most DA(H)e white dwarfs are known to be variable \citep{Elms2023}. Analysis of TESS data for LHS~3250 (including both 20-second and 2-minute cadence observations from several sectors) revealed no significant photometric variations. For WD~J1922+0233, we acquired high-speed photometry of WD~J1922+0233 on UT 2024 June 6 using the APO 3.5\,m telescope with the ARCTIC imager and the BG40 filter. We obtained back-to-back exposures of 25.5\,s over 207\,min under clear skies and $1.0\arcsec$ seeing. To reduce the read-out time, we binned the CCD by $3\times3$, which resulted in a plate scale of $0.34\arcsec$ pixel$^{-1}$. This setup has a read-out time of 4.5\,s, which results in a cadence of $\approx 30$\,s in our light curves. As with LHS~3250, we find no evidence of variability (Figure~\ref{fig:J1922-photometry}) and therefore no additional supporting evidence for this scenario.

An alternative explanation is that temperature inversions occur naturally in the atmospheres of these stars, without any additional heating source. Such inversions can arise in LTE model atmospheres when a change in the dominant absorbing species induces a change in the frequency dependence of the absorption \citep{Dumont1973}. A temperature inversion can then become necessary to satisfy the radiative equilibrium condition. This phenomenon has been observed in some white dwarfs \citep[e.g.,][]{Klein2020} and has been reported for cool white dwarf model atmospheres due to the competition between H$^-$ and CIA opacities \citep{Saumon1994,Bergeron1995}. 

\begin{figure}
\includegraphics[width=\columnwidth]{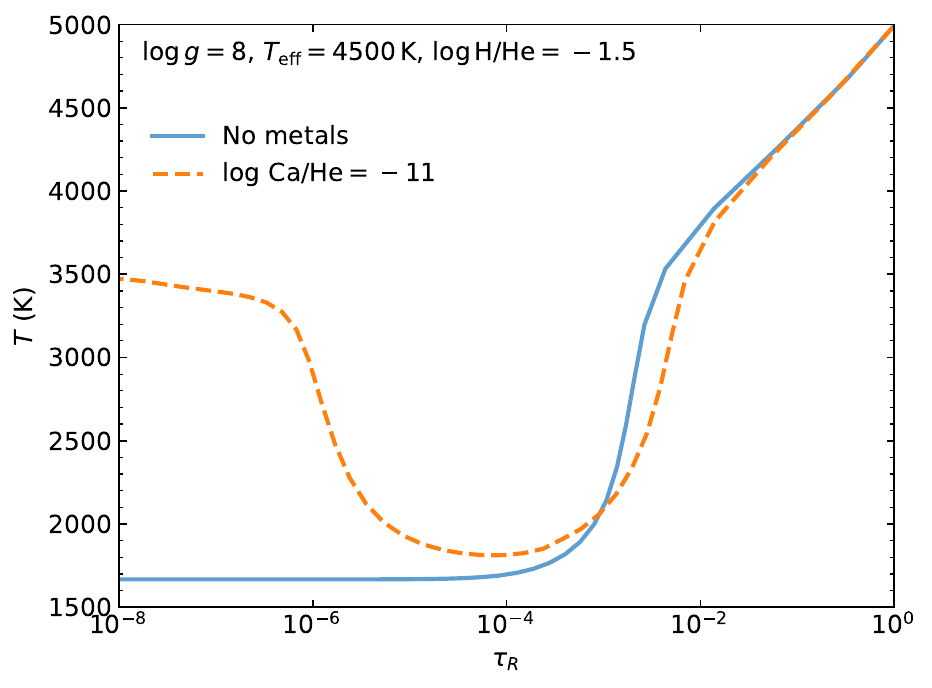}
\caption{Thermal structures of IR-faint model atmospheres at 4500\,K with and without metals. The addition of metals changes the opacity in a way that induces a temperature inversion in the upper atmosphere, while not affecting the photospheric conditions.
\label{fig:metals-temperature-inversion}}
\end{figure}

While our default models of IR-faint white dwarfs do not show the sort of temperature inversions that could explain the observed emission-like features of LHS~3250 and WD~J1922+0233, this does not necessarily rule out this mechanism. The temperature structure in the upper atmosphere is extremely sensitive to the details of different opacity sources, which remain uncertain for these objects (as discussed in Section~\ref{sec:Tinv}). Improvements in our treatment of opacities could potentially lead to models that naturally produce temperature inversions. To illustrate the sensitivity of the temperature structure in the upper layers to small perturbations, Figure~\ref{fig:metals-temperature-inversion} demonstrates how adding a trace amount of metals to the atmosphere of an IR-faint white dwarf is enough to induce a strong temperature inversion. Note however that in this particular case the inversion happens too high above the photosphere to create an emission-like feature in the infrared spectrum. We saw in Figure~\ref{fig:taunu} that for a star with these atmospheric parameters, the $2.4\,\mu$m CIA feature is mostly formed at $\tau_R \sim 10^{-3}$.

A related possibility is that a temperature inversion could be induced by a stratification in the composition of the star's atmosphere \citep{Manseau2016}. In a stratified atmosphere, the hydrogen abundance would be higher in the upper layers due to gravitational settling. However, this seems unlikely for an IR-faint white dwarf. At $T_{\rm eff} = 4500\,$K, $\log g = 8$, and $\log {\rm H/He} = -1.5$, the atmosphere is convective below $\tau_R \simeq 10^{-3}$ and hence completely mixed in that region. One might think that an unmixed chemical stratification is possible above the convective boundary, but overshooting convective plumes should rapidly mix this region as well \citep{Tremblay2013}. This scenario therefore appears to be ruled out.

\section{Keck NIRES Spectroscopy}
\label{sec:keck}
As a complement to our JWST observations, we obtained near-infrared spectroscopy of seven IR-faint white dwarfs with the Near-Infrared Echellette Spectrometer \citep[NIRES,][]{Wilson2004} mounted on the Keck II telescope on UT 2023 September 26. The observed targets were WD J002702.93+\allowbreak055433.39, WD J080440.63+\allowbreak223949.68, WD J172257.78+\allowbreak575250.53, WD J195151.76+\allowbreak402629.07, WD J215008.33$-$\allowbreak043900.36, WD J224206.18+\allowbreak004822.94, and WD J230550.09+\allowbreak392232.87, all known IR-faint white dwarfs \citep{Bergeron2022}. NIRES provides $R = 2700$ spectra over five cross-dispersed orders covering the wavelength range of 0.9$-2.45\,\mu$m. We obtained 5~min long exposures in an ABBA dither pattern along the slit, which was aligned with the parallactic angle. We repeated the observing sequence 3--4 times for each star, resulting in 12 or 16 spectra, and a total on-source integration time of 60 or 80~min. The spectra were extracted using a modified version of the SpeXTool package \citep{Vacca2003,Cushing2004}, with nearby A0~V stars used for telluric correction.

While the noise level of the NIRES spectra is too high to draw conclusions about the peak-CIA region near 2.4$\,\mu$m, they provide valuable information at shorter wavelengths. Our JWST observations of WD~J1922+0233 and LHS~3250 revealed the absence of a 1.2$\,\mu$m feature, which we interpreted as evidence against ultracool temperatures for these objects. The Keck NIRES spectra offer an opportunity to test whether this conclusion extends to a larger sample of IR-faint white dwarfs.

\begin{figure*}
\includegraphics[width=\textwidth]{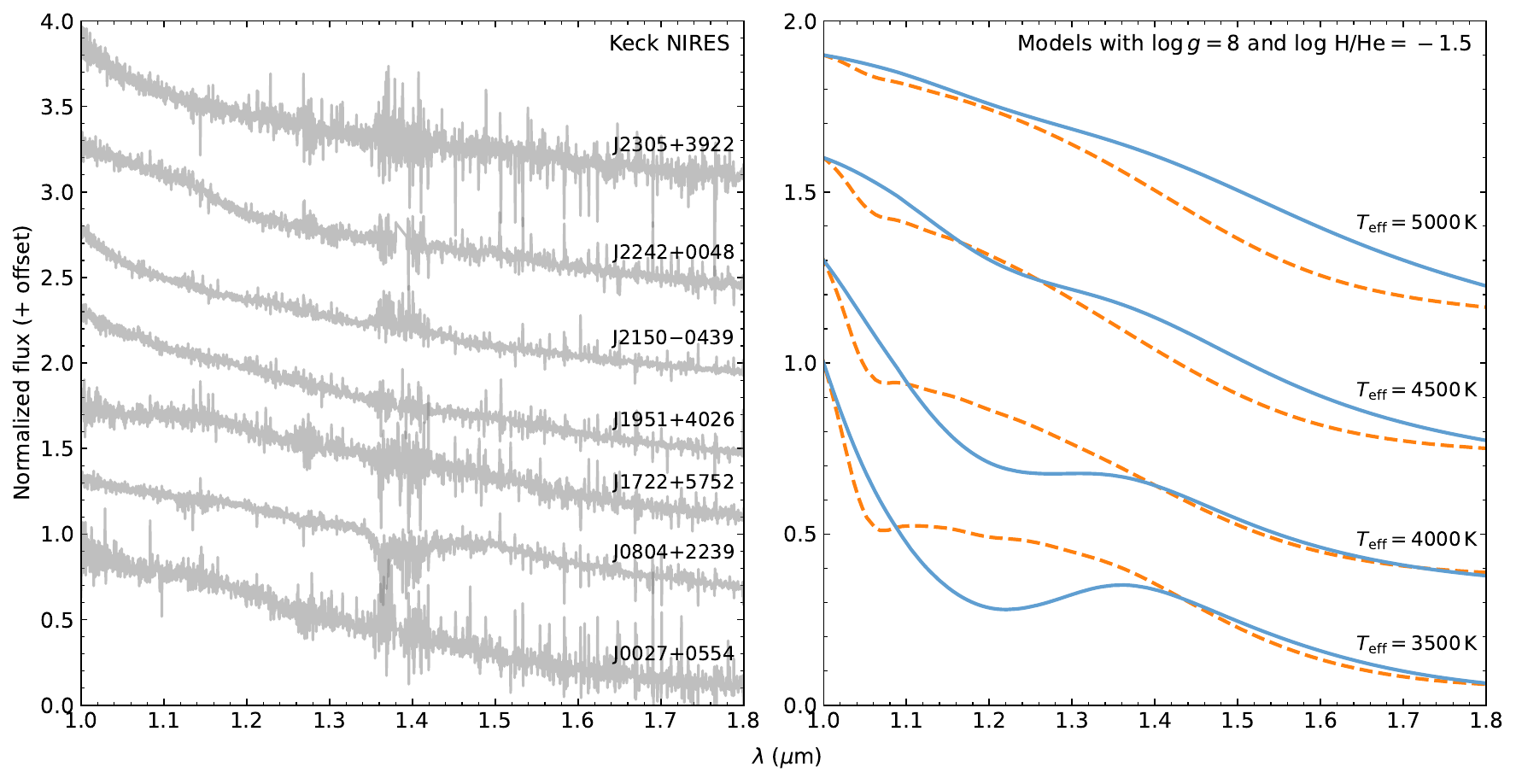}
\caption{\textit{Left}: Keck NIRES spectra of seven IR-faint white dwarfs. The spectra are normalized and vertically offset for clarity. \textit{Right}: Model spectra for white dwarfs with $\log g = 8$ and $\log {\rm H/He} = -1.5$ at various effective temperatures. Solid blue lines represent models using the \citeauthor{Abel2012} H$_2-$He CIA opacities, while dashed orange lines show models using the \citeauthor{Jorgensen2000} opacities. Note the absence of a pronounced molecular hydrogen first overtone band near $1.2\,\mu$m in the observed spectra.
\label{fig:keck}}
\end{figure*}

Figure~\ref{fig:keck} presents the Keck NIRES spectra for our seven additional targets (left panel), along with model predictions for various effective temperatures (right panel). The observed spectra are not compatible with the pronounced 1.2$\,\mu$m features predicted by model atmospheres for ultracool effective temperatures of 4000\,K or below, regardless of whether we use the \citeauthor{Jorgensen2000} or the \citeauthor{Abel2012} CIA opacities (the exact wavelength of the feature differs between both grids of models). This conclusion holds true regardless of the assumed hydrogen abundance, provided it is sufficient to produce significant H$_2$--He CIA. These results further corroborate the ``not-so-cool'' hypothesis of \cite{Bergeron2022}, extending the conclusions drawn from our JWST observations to a broader sample.

\section{Discussion and conclusions}
\label{sec:discussion}
We have presented JWST spectra of three IR-faint white dwarfs, resolving for the first time the precise shape of CIA in those objects. While by far the most striking finding is the emission-like feature detected in the spectrum of WD~J1922+0233, we have seen that the absence of features can sometimes be equally informative. In this last section, we discuss the implications of our findings for each object and outline directions for future research.

LHS~1126 appears to be a distinct case within the IR-faint population. We found that its infrared flux depletion is most likely caused by He--He--He CIA as we do not detect any feature in its JWST spectrum that can be associated with molecular hydrogen. We placed an upper limit of $\log\,{\rm H/He}=-5$ on its photospheric hydrogen abundance, which implies a hydrogen content smaller than $10^{-10}$ of its total mass \citep{Rolland2018}. However, current models appear to be underestimating He--He--He CIA by a factor of $\sim$10, possibly due to uncertainties in the CIA calculations or inaccuracies in the density structure of pure helium atmosphere models. We derived a mass of 0.63\,$M_{\odot}$ for LHS~1126, which is higher than the typical $\simeq$0.55\,$M_{\odot}$ found for cool DQs \citep{Coutu2019,Koester2019,Caron2023}. This is perfectly consistent with the fact that LHS~1126 exhibits unusually weak Swan bands for a DQ white dwarf at its temperature. Indeed, evolutionary models predict that more massive white dwarfs dredge up less carbon from their interiors \citep{Bedard2022}. Finding a higher-than-average mass for a DQ white dwarf with weak carbon features aligns well with these predictions.

For LHS 3250 and WD J1922+0233, the presence of features at 2.4$\,\mu$m, coinciding with peak H$_2$--He CIA absorption, indicates the presence of hydrogen in their atmospheres. While not explicitly discussed earlier, we have ruled out pure hydrogen atmospheres in favor of mixed hydrogen--helium compositions. This is supported by better overall SED fits with mixed models and the stars' position in a sparsely populated region of the color--magnitude diagram (Figure~\ref{fig:HRD}). Since it is expected that even at cool temperatures most white dwarfs have hydrogen-dominated atmospheres, hydrogen-atmosphere white dwarfs must be populating a more crowded region of the color--magnitude diagram (i.e., the cool end of the A/B branch in Figure~\ref{fig:HRD}). The location of LHS~3250 and WD~J1922+0233 in the less populated IR-faint region thus indicates a mixed composition. These objects likely followed the evolutionary pathway described by \cite{Bergeron2022}, where a deepening hydrogen convection zone transforms pure hydrogen atmospheres into mixed hydrogen--helium atmospheres at lower temperatures.

The absence of a 1.2$\,\mu$m CIA feature in the JWST spectra of LHS~3250 and WD~J1922+0233 (and the lack of a pronounced 1.2$\,\mu$m feature in the Keck NIRES spectra of seven additional IR-faint white dwarfs) argues against ultracool temperatures, consistent with the conclusions of \cite{Bergeron2022}. Our analysis also yields relatively high masses for these two stars. A previously unmentioned benefit of these high masses is that they could naturally explain the extreme rarity of metal pollution in IR-faint white dwarfs, with just 1 out of 37 \citep{Bergeron2022} compared to $\sim$30\% of helium-atmosphere white dwarfs in the 4000--5000\,K range. This is because massive white dwarfs are generally much less likely to be metal-polluted \citep{Koester2014}. However, the narrowness of WD~J1922+0233's Na absorption line in the optical remains unexplained.

The fact that we observe an emission-like feature at 2.4$\,\mu$m in two out of two IR-faint white dwarfs with hydrogen--helium atmospheres suggests that such features may be ubiquitous in this population. This previously unrecognized phenomenon may explain the persistent difficulties in fitting IR-faint white dwarf SEDs over the past 25 years. Our analysis suggests that the emission-like features observed in IR-faint white dwarfs could be explained by a temperature inversion above the photosphere. However, without a good predictive model to explain this behavior, the precise parameters of IR-faint white dwarfs remain uncertain. Unlike most stellar atmospheres where upper atmospheric effects primarily influence spectral line cores, the situation is different for IR-faint white dwarfs, since the infrared continuum is strongly affected by the CIA acting well above the photosphere. This implies that even photometric fits are sensitive to these temperature inversions. Until we develop an understanding of the mechanisms causing this temperature inversion and incorporate it into our models, precise characterization of IR-faint white dwarfs with mixed atmospheres will remain challenging.

We end by highlighting several research directions that could help further elucidate the nature of IR-faint white dwarfs. First, spectropolarimetric observations of additional IR-faint white dwarfs could reveal magnetic fields, thereby testing the idea that these white dwarfs host magnetic fields that support a chromosphere responsible for the observed emission-like features. Second, theoretical and laboratory studies of He$^-$ free-free absorption, H$^-$ bound-free absorption, Ly$\alpha$ broadening, and CIA under conditions relevant to cool white dwarf atmospheres would improve our understanding of these objects and cool white dwarfs in general. Last but not least, additional JWST observations of IR-faint white dwarfs are crucial to determine whether the $2.4\,\mu$m emission-like feature is ubiquitous in this population. The diverse spectra observed in the three stars studied here further underscore the need for additional observations. Each object presents unique characteristics, suggesting that a larger sample may reveal even more unexpected features.

\section*{Acknowledgements}
SB thanks Pierre Bergeron for insightful discussions on model atmospheres and René Doyon for encouraging the application for time to observe IR-faint white dwarfs with JWST. The authors thank the anonymous referee for their useful comments. SB acknowledges the support of the Canadian Space Agency (CSA) [23JWGO2A10] and of the Canadian Institute for Theoretical Astrophysics (CITA). MK acknowledges support by the NSF under grant AST-2205736 and NASA under grants No.
80NSSC22K0479, 80NSSC24K0380, and 80NSSC24K0436.

This work is based on observations made with the NASA/ESA/CSA James Webb Space Telescope. The data were obtained from the Mikulski Archive for Space Telescopes at the Space Telescope Science Institute, which is operated by the Association of Universities for Research in Astronomy, Inc., under NASA contract NAS 5-03127 for JWST. These observations are associated with program \#3168. Support for this program was provided through a grant from the STScI under NASA contract NAS5-03127.

The Apache Point Observatory 3.5-meter telescope is owned and operated by the Astrophysical Research Consortium.

This work was supported by a NASA Keck PI Data Award, administered by the NASA Exoplanet Science Institute. Data presented herein were obtained at the W. M. Keck Observatory from telescope time allocated to the National Aeronautics and Space Administration through the agency's scientific partnership with the California Institute of Technology and the University of California. The Observatory was made possible by the generous financial support of the W. M. Keck Foundation.

The authors wish to recognize and acknowledge the very significant cultural role and reverence that the summit of Maunakea has always had within the indigenous Hawaiian community. We are most fortunate to have the opportunity to conduct observations from this mountain.

This work has made use of data from the European Space Agency (ESA) mission
{\it Gaia} (\url{https://www.cosmos.esa.int/gaia}), processed by the {\it Gaia}
Data Processing and Analysis Consortium (DPAC,
\url{https://www.cosmos.esa.int/web/gaia/dpac/consortium}). Funding for the DPAC
has been provided by national institutions, in particular the institutions
participating in the {\it Gaia} Multilateral Agreement.

This work has made use of the Montreal White Dwarf Database \citep{Dufour2017}.

\bibliography{references}{}
\bibliographystyle{aasjournal}



\end{document}